\documentclass[11pt,a4paper]{article}
\pdfoutput=1
\usepackage{jheppub}

\usepackage{multirow, graphicx,amssymb,url,mathrsfs,amsmath}
\usepackage{wrapfig,boxedminipage,setspace,subfigure,epsfig}
\usepackage{amsxtra,amstext,latexsym,dsfont,amsfonts}
\usepackage{color,eucal}
\usepackage[dvipsnames]{xcolor}
\usepackage{float}
\usepackage{slashed,comment}
\usepackage{kotex}
\usepackage{tikz}
\usetikzlibrary{calc,patterns,angles,quotes}
\usetikzlibrary{decorations.pathreplacing,decorations.markings,snakes}
\usepackage{tabularx, array}
\usepackage{mdframed,mathtools}
\newcolumntype{L}[1]{>{\raggedright\arraybackslash}p{#1}}
\newcolumntype{C}[1]{>{\centering\arraybackslash}p{#1}}
\newcolumntype{R}[1]{>{\raggedleft\arraybackslash}p{#1}}

\newcommand{\be}{\begin{equation}}
\newcommand{\ee}{\end{equation}}
\newcommand{\bea}{\begin{eqnarray}}
\newcommand{\eea}{\end{eqnarray}}

\title{A geodesic distance interpretation of Lanczos coefficients}

\author[a,b]{Le-Chen Qu}
\emailAdd{lechen.qu@ift.csic.es}

\preprint{\texttt{IFT-UAM/CSIC-26-111}}

\affiliation[a]{Instituto de F\'isica Te\'orica UAM/CSIC, Calle Nicol\'as Cabrera 13-15, 28049 Madrid, Spain}
\affiliation[b]{Departamento de F\'isica Te\'orica, Universidad Aut{\'o}noma de Madrid, 28049 Madrid, Spain}

\abstract{We study the spread complexity of the infinite-temperature
thermofield double state in large-$N$ random matrix theory and investigate the
physical meaning of its Lanczos coefficients $b_n$. The leading density of states
defines a family of orthogonal polynomials whose recursion coefficients
coincide with the Lanczos coefficients of the associated Krylov dynamics.
Starting from the Coulomb-gas equation, we derive a pair of
nonlinear relations for these coefficients, which we call the Lanczos
equations. For normalizable one-cut ensembles, the Lanczos coefficients
approach constants at large Krylov index, rendering the asymptotic Krylov
chain translationally invariant and the late-time growth of spread complexity
linear, with the growth rate given by the spectral average of the group
velocity obtained from a Bloch-wave analysis. For an even potential, we further show that the
difference between consecutive squared Lanczos coefficients,
$b_{n+1}^2-b_n^2$, is the generating function for two-legged maps whose marked
legs are separated by the exact geodesic distance $n$. Consequently, the spread complexity
acceleration operator equals twice the geodesic generating operator. We
illustrate these results in the quartic matrix model, where the expansion in
the quartic coupling counts tetravalent planar maps, and in the double-scaled
Sachdev--Ye--Kitaev model, where the auxiliary Hilbert space can be
identified with the physical chord Hilbert space.}

\begin{document}
\maketitle

%

\section{Introduction}
A simple initial state evolving under a chaotic Hamiltonian typically develops into a complex state that spreads throughout Hilbert space. The Lanczos algorithm
maps this abstract spreading onto motion along a one-dimensional chain, whose sites correspond to the Krylov basis vectors \cite{Parker:2018yvk}. The spread
complexity is the average position of the evolving state along this chain and
thus quantifies how far it has propagated from the initial state~\cite{Balasubramanian:2022tpr}.
For comprehensive reviews, see Refs.~\cite{Nandy:2024htc,Baiguera:2025dkc,Rabinovici:2025otw}. Although spread complexity has
been studied extensively \cite{Barbon:2019wsy,Avdoshkin:2019trj,Rabinovici:2020ryf,Jian:2020qpp,Jaiswal:2020snm,Dymarsky:2021bjq,Hornedal:2022pkc,He:2022ryk,Caputa:2023vyr,Erdmenger:2023wjg,Craps:2023ivc,Huh:2023jxt,He:2024hkw,He:2024xjp,Caputa:2024vrn,Baggioli:2024wbz,Craps:2024suj,Huh:2024ytz,Caputa:2024sux,Zhai:2024tkz,Nandy:2024mml,Li:2024ljz,Balasubramanian:2024ghv,Bhattacharya:2024szw,Bhattacharya:2024hto,Bhattacharya:2024uxx,Aguilar-Gutierrez:2024nau,Baggioli:2025knt,Craps:2025kub,Evnin:2025cfx,Fu:2025kkh,He:2025guu,Zhai:2025abc,Caputa:2025dep,Caputa:2025ozd,Caputa:2025mii,Miyaji:2025yvm,Takahashi:2025iol,Miyaji:2025ucp,Demulder:2025uda,Imani:2025etp,Alishahiha:2026fnu,Chowdhury:2026fjb,Li:2026jxx,DeRo:2026mlc,Nunez:2026kwr,Das:2026gko,Bhattacharyya:2026zpu,Balasubramanian:2026klv,Caputa:2026ldd}, much less attention has been devoted to the physical
meaning of the Lanczos coefficients themselves. In the conventional Krylov chain
description, these coefficients appear merely as hopping amplitudes and on-site
energies, and no comparably direct physical interpretation has yet been
established for them.

From a complementary mathematical perspective, orthogonal polynomials can be
reinterpreted as Krylov polynomials because their three-term recurrence
relation has precisely the form of the Lanczos recursion, allowing the
recursion coefficients to be identified with the Lanczos
coefficients~\cite{Muck:2022xfc, Kar:2021nbm, Muck:2024fpb, Adhikari:2025vdl, Alishahiha:2024vbf,Balasubramanian:2025xkj,Lunt:2025dcc,Balasubramanian:2022dnj,Qu:2025lgo,Murugan:2026rfa,Pedraza:2026zji}. This connection is especially natural in
random matrix theory, where orthogonal polynomials play a central role and
their recursion coefficients obey the well-known discrete string
equations~\cite{Brezin:1977sv,Bessis:1980ss,Itzykson:1979fi}. Nevertheless,
the polynomial basis is usually constructed in an auxiliary Hilbert space and does not, by itself, provide a representation of the physical
Hilbert space of the underlying theory. Without an additional physical
identification, the evolution generated in this auxiliary space therefore has
no immediate physical meaning. 

In this work, we address these questions by studying the one-sided evolution
of the infinite-temperature thermofield double state in large-$N$ random
matrix ensembles. We show that the resulting dynamics can be formulated in
terms of polynomials orthogonal with respect to the leading density of states.
Starting from the Coulomb-gas equation, we then derive a set of nonlinear
relations satisfied by the associated Lanczos coefficients. These
relations resemble the discrete string equations but govern the Lanczos coefficients associated with the density of states; we therefore refer to them
as the Lanczos equations. For a normalizable one-cut random matrix ensemble,
the Lanczos coefficients $a_n$ and $b_n$ approach constants at large Krylov
index, with their limiting values determined by the Lanczos equations. The
Krylov chain therefore becomes asymptotically translationally invariant,
causing the spread complexity acceleration to vanish and $C(t)$ to grow
linearly at late times. A Bloch-wave analysis of this asymptotic chain further
yields an analytic expression for the growth rate as the spectral average of
the corresponding group velocity. More unexpectedly, for an even potential we
identify the difference between consecutive squared Lanczos coefficients,
$b_{n+1}^2-b_n^2$, as the generating function for planar maps whose two marked
legs are separated by the exact geodesic distance $n$
\cite{Bouttier:2003dh,DiFrancesco:2004qj,di2005geodesic}. It follows that the
spread complexity acceleration operator equals twice the geodesic generating
operator, and hence that the acceleration of $C(t)$ is twice the expectation
value of this operator in the evolving state.

We illustrate this framework using the quartic matrix model and the
double-scaled Sachdev--Ye--Kitaev (DSSYK) model
\cite{Cotler:2016fpe,Garcia-Garcia:2018fns,Berkooz:2018jqr}. In the quartic
model, we show that the coefficient of $g^k$ in
$b_{n+1}^2-b_n^2$ counts planar maps with two legs, $k$ tetravalent internal
vertices and geodesic distance $n$. In the DSSYK model, the auxiliary Hilbert
space constructed from the orthogonal polynomials can be identified with the
chord Hilbert space, which provides a concrete representation of the physical
Hilbert space of the theory~\cite{Lin:2022rbf,Rabinovici:2023yex,Balasubramanian:2024lqk}.

The remainder of this paper is organized as follows. In
section~\ref{Hamiltonian}, we formulate the dynamics of a single Hamiltonian
and then extend the construction to large-$N$ random matrix ensembles.
Section~\ref{sec:orthogonal-polynomials} develops the orthogonal polynomial
description of the resulting Krylov dynamics. In
section~\ref{sec:lanczos-equations}, we derive the Lanczos equations, compare
them with the discrete string equations, and study the large-$n$ behavior of Lanczos coefficients. Section~\ref{sec:geodesic-distance} establishes the
combinatorial interpretation of the Lanczos coefficients in terms of planar
maps and relates spread complexity acceleration to geodesic distance
enumeration. We then apply the general framework to the quartic matrix model
in section~\ref{sec:quartic-potential} and to the DSSYK model in
section~\ref{sec:dssyk}. Finally, section~\ref{Conclusion} summarizes our
results and discusses possible directions for future work.

\section{Hamiltonian dynamics and random matrix ensembles}\label{Hamiltonian}
Let $H$ be a Hamiltonian acting on an $N$-dimensional Hilbert space $\mathcal{H}$, with eigenstates $|E_i\rangle$ and eigenvalues $E_i$. On the doubled Hilbert space $\mathcal{H}_L\otimes\mathcal{H}_R$, the infinite-temperature thermofield double (TFD) state is \cite{Balasubramanian:2022tpr,Rabinovici:2023yex}
\begin{equation}
|\mathrm{TFD}_\infty\rangle
=\frac{1}{\sqrt{N}}
\sum_{i=1}^{N}
|E_i\rangle_L|E_i\rangle_R.
\end{equation}
Here $H_L=H\otimes\mathds{1}$ and $H_R=\mathds{1}\otimes H$ act on the left and right copies of $\mathcal{H}$, respectively. Since $(H_L-H_R)|\mathrm{TFD}_\infty\rangle=0$, the TFD state is invariant under the evolution generated by $H_L-H_R$. By contrast, evolution generated by $H_L$ alone is nontrivial and gives
\begin{equation}\label{evolu}
e^{-iH_Lt}|\mathrm{TFD}_\infty\rangle
=
\frac{1}{\sqrt{N}}
\sum_{i=1}^{N}
e^{-iE_it}|E_i\rangle_L|E_i\rangle_R.
\end{equation}
Both the TFD state and its time evolution lie in the diagonal subspace spanned by the paired energy eigenstates in Eq.~\eqref{evolu}. Identifying each paired state with the corresponding state $|E_i\rangle$ maps this dynamics to evolution in a single copy of $\mathcal{H}$ under $H$, with initial state
\begin{equation}\label{evoluti1}
|\phi(0)\rangle
=
\frac{1}{\sqrt{N}}
\sum_{i=1}^{N}
|E_i\rangle.
\end{equation}
The corresponding time-evolved state is
\begin{equation}\label{phievolution}
|\phi(t)\rangle
=
e^{-iHt}|\phi(0)\rangle
=
\frac{1}{\sqrt{N}}
\sum_{i=1}^{N}
e^{-iE_it}|E_i\rangle.
\end{equation}
This evolution can be characterized by the survival amplitude
\begin{equation}\label{surviampli}
S(t)
=
\langle \phi(t)|\phi(0)\rangle
=
\langle \phi(0)|e^{iHt}|\phi(0)\rangle
=
\frac{1}{N}\operatorname{Tr}\!\left(e^{iHt}\right)
=
\sum_{n=0}^{\infty}m_n\frac{(it)^n}{n!},
\end{equation}
where the moments of the Hamiltonian are
\begin{equation}
m_n=\frac{1}{N}\operatorname{Tr}\!\left(H^n\right).
\end{equation}
A Laplace-type transform of $S(t)$ gives the Green's function
\begin{equation}
G(z)
=
i\int_{0}^{\infty}dt\,e^{-izt}S(t)
=
\sum_{n=0}^{\infty}\frac{m_n}{z^{n+1}}
=
\frac{1}{N}\operatorname{Tr}\!\left(\frac{1}{z-H}\right)
=
\frac{1}{N}\sum_{i=1}^{N}\frac{1}{z-E_i}.
\end{equation}
In the matrix integral literature, $G(z)$ is known as the normalized matrix resolvent. For finite $N$, it is a meromorphic function with simple poles at the eigenvalues $E_i$. The normalized density of states is recovered from its boundary values on the real axis through the Sokhotski--Plemelj formula
\begin{equation}
\lim_{\epsilon\to0^+}
\left[G(E+i\epsilon)-G(E-i\epsilon)\right]
=
-2\pi i\,\rho(E),
\end{equation}
where the equality is understood in the sense of distributions, and the normalized density of states is
\begin{equation}
\rho(E)
=
\frac{1}{N}\sum_{i=1}^{N}
\delta(E-E_i).
\end{equation}
It follows immediately that the moments can equivalently be written as
\begin{equation}\label{momentsandrho}
\int dE\,E^n\rho(E)
=
\frac{1}{N}
\sum_{i=1}^{N}
E_i^n
=
\frac{1}{N}\operatorname{Tr}\!\left(H^n\right)
=m_n.
\end{equation}
We now pass from the dynamics generated by a fixed Hamiltonian to its ensemble-averaged counterpart by treating $H$ as an $N\times N$ Hermitian random matrix distributed according to the probability measure \cite{BESSIS1980109,Ginsparg:1991bi,Eynard:2015aea,bleher2011lectures,Ginsparg:1993is,livan2018introduction}
\begin{equation}
    d\mu_N(H)
    =
    \frac{1}{Z_N}
    e^{-N\operatorname{Tr}V(H)}\,dH.
\end{equation}
where the potential $V(\lambda)$ is assumed to be a polynomial satisfying
\begin{equation}\label{boundarycondition}
    \lim_{\lambda\to\pm\infty} \bigl( V(\lambda) - \log(\lambda^2+1) \bigr) = +\infty,
\end{equation}
so that the measure is normalizable. The Lebesgue measure on the space of Hermitian matrices is
\begin{equation}
    dH
    =
    \prod_{j=1}^{N}dH_{jj}
    \prod_{1\leq j<k\leq N}
    d\,\operatorname{Re}H_{jk}\,
    d\,\operatorname{Im}H_{jk},
\end{equation}
The normalization factor $Z_N$ is the partition function
\begin{equation}
    Z_N
    =
    \int
    e^{-N\operatorname{Tr}V(H)}\,dH.
\end{equation}
With this normalization, the ensemble average of any integrable observable $f(H)$ is
\begin{equation}
    \langle f(H)\rangle
    =
    \int f(H)\,d\mu_N(H)
    =
    \frac{1}{Z_N}
    \int
    f(H)e^{-N\operatorname{Tr}V(H)}\,dH.
\end{equation}
Since both $dH$ and $\operatorname{Tr}V(H)$ are invariant under unitary conjugation, the measure $d\mu_N(H)$ is invariant under
\begin{equation}
    H\longrightarrow UHU^\dagger,
    \qquad U\in U(N).
\end{equation}
The Weyl integration formula then gives the joint eigenvalue measure
\begin{equation}\label{jointprobabilitydensity}
    d\mu_N(E_1,\ldots,E_N)
    =
    \frac{1}{\widetilde{Z}_N}
    \prod_{1\leq j<k\leq N}(E_j-E_k)^2
    \exp\!\left[-N\sum_{i=1}^{N}V(E_i)\right]
    \prod_{i=1}^{N}dE_i,
\end{equation}
where the normalization constant is
\begin{equation}
    \widetilde{Z}_N
    =
    \int
    \prod_{1\leq j<k\leq N}(E_j-E_k)^2
    \exp\!\left[-N\sum_{i=1}^{N}V(E_i)\right]
    \prod_{i=1}^{N}dE_i.
\end{equation}
In the large-$N$ limit, the eigenvalues become densely spaced, and the density of states $\rho(E)$ approaches a smooth deterministic profile on macroscopic scales. Fluctuations about this profile are suppressed, a property known as self-averaging. We define the leading density of states by \cite{Saad:2019lba}
\begin{equation}\label{leadingdenstates}
    \rho_0(E)
    =
    \lim_{N\to\infty}\langle\rho(E)\rangle.
\end{equation}
The leading density $\rho_0(E)$ is determined by the large-$N$ saddle point of the joint measure in Eq.~\eqref{jointprobabilitydensity}. The effective potential experienced by a single eigenvalue contains the confining potential and the logarithmic repulsion induced by the Vandermonde factor,
\begin{equation}\label{veffvander}
    V_{\mathrm{eff}}(E_j)
    =
    N V(E_j)
    -
    \sum_{i\neq j}
    \log\!\left[(E_i-E_j)^2\right].
\end{equation}
In the continuum approximation,
\begin{equation}
    \sum_{i\ne j}
    \log\!\left[(E_i-E_j)^2\right]
   =
    N\int dE\,\rho_0(E)
    \log\!\left[(E-E_j)^2\right].
\end{equation}
Imposing the stationarity condition $V_{\mathrm{eff}}'(E_j)=0$ then gives the Coulomb-gas equation
\begin{equation}\label{coulombgas-saddle}
    \frac{1}{2}V'(E_j)
    =
   \!\int dE\,
    \frac{\rho_0(E)}{E_j-E}.
\end{equation}
The integral is understood in the Cauchy principal value sense, which is the continuum counterpart of omitting the $i=j$ term in Eq.~\eqref{veffvander}. Eq.~\eqref{momentsandrho} then gives the large-$N$ moments\footnote{Here, the ensemble average is performed at the level of the moments, leading to the annealed complexity \cite{Parker:2018yvk}, rather than the quenched complexity obtained by averaging the complexity itself, as in Ref.~\cite{Balasubramanian:2022tpr}. As emphasized in Ref.~\cite{Caputa:2024vrn}, these two prescriptions generally lead to different results. Clarifying the precise relation between them is an interesting direction for future work.},
\begin{equation}\label{largeoments}
    \lim_{N\to\infty}\langle m_n\rangle
    =
    \int dE\,E^n\rho_0(E).
\end{equation}
These moments determine the large-$N$ averaged resolvent and its Jacobi continued-fraction representation \cite{Parker:2018yvk},
\begin{equation}\label{greenexpansin}
    \lim_{N\to\infty}\langle G(z)\rangle
    =
    \lim_{N\to\infty}
    \sum_{n=0}^{\infty}
    \frac{\langle m_n\rangle}{z^{n+1}}
    =
    \frac{1}{
    z-a_0-
    \dfrac{b_1^2}{
    z-a_1-
    \dfrac{b_2^2}{
    z-a_2-\cdots}}}.
\end{equation}
Here $a_n$ and $b_n$ are the Lanczos coefficients associated with the large-$N$ moments. They can be computed using the recursive algorithm \cite{viswanath1994recursion,bhattacharjee2023operator}
\begin{equation}\label{recursivealgorithm}
\begin{aligned}
M_k^{(0)}
&=
(-1)^k \lim_{N\to\infty}\langle m_k \rangle,
\qquad
L_k^{(0)}
=
(-1)^{k+1} \lim_{N\to\infty}\langle m_{k+1} \rangle,
\\
M_k^{(n)}
&=
L_k^{(n-1)}
-
L_{n-1}^{(n-1)}
\frac{M_k^{(n-1)}}{M_{n-1}^{(n-1)}}, \qquad
L_k^{(n)}
=
\frac{M_{k+1}^{(n)}}{M_n^{(n)}}
-
\frac{M_k^{(n-1)}}{M_{n-1}^{(n-1)}},
\qquad
k\ge n,
\\
b_n
&=
\sqrt{M_n^{(n)}},
\qquad
a_n
=
-
L_n^{(n)}.
\end{aligned}
\end{equation}
In the next section, we relate these Lanczos coefficients to the recursion coefficients of orthogonal polynomials.

\section{Orthogonal polynomial approach}\label{sec:orthogonal-polynomials}
Let $d\mu(\lambda)=\rho_0(\lambda)d\lambda$ denote the normalized measure associated with the leading density of states, so that $\int d\mu(\lambda)=1$. The corresponding monic orthogonal polynomials are defined by
\begin{equation}\label{innerproduct-rho0}
    P_n(\lambda)
    =
    \lambda^n+\text{(lower degree terms)},
    \qquad
    \int
    P_n(\lambda)P_m(\lambda)\,d\mu(\lambda)
    =
    h_n\delta_{mn}.
\end{equation}
Here $h_n>0$ is the squared norm of $P_n$, with $P_0=1$ and $h_0=1$. Introducing the Vandermonde determinant $\Delta(\boldsymbol{\lambda})=\prod_{1\leq i<j\leq n}(\lambda_i-\lambda_j)$, Heine's formula gives
\begin{equation}\label{Heineformula}
P_n(\lambda)
=
\frac{
\int
\prod_{i=1}^{n}(\lambda-\lambda_i)
\Delta(\boldsymbol{\lambda})^2
\prod_{i=1}^{n}d\mu(\lambda_i)
}{
\int
\Delta(\boldsymbol{\lambda})^2
\prod_{i=1}^{n}d\mu(\lambda_i)
}.
\end{equation}
With $P_{-1}=0$ and $P_0=1$, the monic orthogonal polynomials satisfy the three-term recurrence relation
\begin{equation}\label{recursionrelation-rho0}
    \lambda\, P_n(\lambda)
    = P_{n+1}(\lambda)
      + S_n\, P_n(\lambda)
      + R_n\, P_{n-1}(\lambda),
    \qquad n\geq0,
\end{equation}
where
\begin{equation}
    S_n
    =
    \frac{1}{h_n}
    \int
    \lambda P_n(\lambda)^2\,d\mu(\lambda),
    \qquad
    R_n
    =
    \frac{h_n}{h_{n-1}}>0
    \quad (n\geq1).
\end{equation}
This truncation to three terms follows directly from the orthogonality condition in Eq.~\eqref{innerproduct-rho0}, which ensures that $\lambda P_n$ has components only along $P_{n+1}$, $P_n$, and $P_{n-1}$. It is useful to introduce an auxiliary Hilbert space spanned by the normalized states
\begin{equation}\label{auxiliaryHilbertspace}
    |n\rangle = \frac{P_n(\lambda)}{\sqrt{h_n}}, \qquad n = 0,1,2,\ldots ,
\end{equation}
which form an orthonormal basis,
\begin{equation}
    \langle n|m\rangle
    =
    \frac{1}{\sqrt{h_nh_m}}
    \int
    P_n(\lambda)P_m(\lambda)\,d\mu(\lambda)
    =
    \delta_{nm}.
\end{equation}
Within the auxiliary Hilbert space, the recursion relation in Eq.~\eqref{recursionrelation-rho0} becomes
\begin{equation}\label{compactrecursionrelation-rho0}
    \hat{\lambda}\, |n\rangle
    = \sqrt{R_{n+1}}\, |n+1\rangle
      + S_n\, |n\rangle
      + \sqrt{R_n}\, |n-1\rangle.
\end{equation}
Thus, $\hat{\lambda}$ takes the tridiagonal Jacobi form characteristic of a Hamiltonian represented in a Krylov basis. We consider the evolution generated by $\hat{\lambda}$ from the initial state $|0\rangle$ \cite{Qu:2025lgo},
\begin{equation}\label{krylovdyna}
|\psi(t)\rangle
= e^{-i\hat{\lambda} t} |0\rangle
= \sum_{n=0}^{\infty} \frac{(-i t)^{n}}{n!}\, \hat{\lambda}^{\,n} |0\rangle
= \sum_{n=0}^{\infty} \psi_n(t)\, |n\rangle,
\end{equation}
where $\psi_n(t)=\langle n|\psi(t)\rangle$. With the convention adopted in Eq.~\eqref{surviampli}, the corresponding survival amplitude is
\begin{equation}\label{survivalamplitude}
\begin{aligned}
S_0(t)
&=
\langle\psi(t)|0\rangle
=
\langle0|e^{i\hat{\lambda}t}|0\rangle
=
\int e^{i\lambda t}\,d\mu(\lambda)
\\
&=
\sum_{k=0}^{\infty}
\frac{(it)^k}{k!}
\int \lambda^k\rho_0(\lambda)\,d\lambda
=
\lim_{N\to\infty}\langle S(t)\rangle,
\end{aligned}
\end{equation}
which is precisely the ensemble-averaged survival amplitude in Eq.~\eqref{surviampli}. Since the survival amplitude uniquely determines the Krylov dynamics, the states
\(
|n\rangle=P_n(\lambda)/\sqrt{h_n}
\)
form a natural Krylov basis for the large-$N$ ensemble-averaged dynamics
\(
|\phi(t)\rangle=e^{-iHt}|\phi(0)\rangle
\)
in random matrix theory. Comparing Eq.~\eqref{compactrecursionrelation-rho0} with the Jacobi continued fraction in Eq.~\eqref{greenexpansin} gives
\begin{equation}\label{coresplanrec}
    a_n=S_n,
    \qquad
    b_n^2=R_n.
\end{equation}
The Krylov amplitudes consequently obey the discrete Schr\"odinger equation \cite{Balasubramanian:2022tpr}
\begin{equation}\label{psieq}
i\,\partial_t \psi_n(t)
= b_{n+1}\,\psi_{n+1}(t)
  + a_n\,\psi_n(t)
  + b_n\,\psi_{n-1}(t),
\qquad
\psi_n(0)=\delta_{n0},
\end{equation}
with the Krylov amplitudes expressed in terms of the monic orthogonal polynomials as
\begin{equation}
\psi_n(t)
=\frac{1}{\sqrt{h_n}}\langle 0 | P_n(\hat{\lambda}) e^{-i\hat{\lambda} t} | 0 \rangle=
\frac{1}{\sqrt{h_n}}
\int d\mu(\lambda)\,
P_n(\lambda)e^{-i\lambda t}.
\end{equation}
Finally, the spread complexity is defined by \cite{Balasubramanian:2022tpr}
\begin{equation}\label{spreadcomplexity}
C(t)
= \sum_{n=0}^{\infty}n\,|\psi_n(t)|^2
= \sum_{n=0}^{\infty}\frac{n}{h_n}
  \int
  d\mu(\lambda)\,d\mu(\lambda')\,
  P_n(\lambda)P_n(\lambda')
  e^{-i(\lambda-\lambda')t}.
\end{equation}
It is the mean position of the state along the semi-infinite Krylov chain and therefore quantifies its spreading away from the initial state.

\section{Discrete string equations and Lanczos equations}\label{sec:lanczos-equations}
The polynomials $P_n(\lambda)$ introduced above should be distinguished from the standard monic orthogonal polynomials
\(
p_n(\lambda)=\lambda^n+\text{(lower degree terms)},
\)
used in random matrix theory. The latter are orthogonal with respect to the measure $d\nu(\lambda)=e^{-NV(\lambda)}d\lambda$ \cite{BESSIS1980109,Ginsparg:1991bi,Eynard:2015aea,bleher2011lectures,Ginsparg:1993is,livan2018introduction},
\begin{equation}\label{innerproduct}
    \int
    p_n(\lambda)p_m(\lambda)\,d\nu(\lambda)
    =
    H_n\delta_{mn}.
\end{equation}
Here $H_n>0$ is the squared norm of $p_n$. Since $p_0=1$, one has $H_0=\int d\nu(\lambda)$. The monic polynomials satisfy the three-term recurrence relation
\begin{equation}\label{recursionrelation-matrixmodel}
    \lambda\, p_n(\lambda)
    = p_{n+1}(\lambda)
      + s_n\, p_n(\lambda)
      + r_n\, p_{n-1}(\lambda),
\end{equation}
with $p_{-1}=0$, $p_0=1$, and $r_n=H_n/H_{n-1}>0$ for $n\geq1$. Monicity and orthogonality imply
\begin{equation}\label{twoidentitiess}
\begin{aligned}
    \int d\nu(\lambda)\,
    p_n(\lambda)\frac{d}{d\lambda}p_n(\lambda)
    &=0,
    \\
    \int d\nu(\lambda)\,
    p_{n-1}(\lambda)\frac{d}{d\lambda}p_n(\lambda)
    &=nH_{n-1}.
\end{aligned}
\end{equation}
The normalized polynomials define states in the auxiliary Hilbert space,
\begin{equation}
    |n\rangle
    =
    \frac{p_n(\lambda)}{\sqrt{H_n}},
    \qquad n=0,1,2,\ldots,
\end{equation}
and form an orthonormal basis,
\begin{equation}\label{orthonormalbasis}
    \langle n|m\rangle
    =
    \delta_{nm}.
\end{equation}
In this basis, the recursion relation in Eq.~\eqref{recursionrelation-matrixmodel} takes the Jacobi form
\begin{equation}\label{compactrecursionrelation}
    \hat{\lambda}\, |n\rangle
    = \sqrt{r_{n+1}}\, |n+1\rangle
      + s_n\, |n\rangle
      + \sqrt{r_n}\, |n-1\rangle.
\end{equation}
For a confining potential, the boundary terms vanish. Integration by parts, together with Eq.~\eqref{twoidentitiess}, then yields the discrete string equations
\begin{equation}\label{VSRREEQ}
\begin{aligned}
\langle n|V'(\hat{\lambda})|n\rangle
&=0,
\\
\sqrt{r_n}\,\langle n-1|V'(\hat{\lambda})|n\rangle
&=\frac{n}{N}.
\end{aligned}
\end{equation}
Given initial conditions, the recursion coefficients $r_n$ and $s_n$ can be determined from the discrete string equations~\eqref{VSRREEQ}. We now show that the recursion coefficients $R_n$ and $S_n$ satisfy analogous relations, which we refer to as the Lanczos equations. Recall that the leading density of states obeys the Coulomb-gas equation
\begin{equation}\label{coulombgas2}
\frac{1}{2}V'(\lambda)
=
\!\int dE\,
\frac{\rho_0(E)}{\lambda-E}.
\end{equation}
For any polynomial $f(\lambda)$, multiplying Eq.~\eqref{coulombgas2} by $2\rho_0(\lambda)f(\lambda)$, integrating over $\lambda$, and symmetrizing under $\lambda\leftrightarrow E$ yields\footnote{The difference quotient has the finite coincidence limit
$\lim_{\lambda\to E}[f(\lambda)-f(E)]/(\lambda-E)=f'(E)$. The apparent singularity at $\lambda=E$ is therefore removable, and the integral in Eq.~\eqref{eq:symmetrized-saddle} is an ordinary rather than a principal-value integral.}
\begin{equation}\label{eq:symmetrized-saddle}
\int d\lambda\,
V'(\lambda)\rho_0(\lambda)f(\lambda)
=
\int d\lambda\,dE\,
\rho_0(\lambda)\rho_0(E)
\frac{f(\lambda)-f(E)}{\lambda-E}.
\end{equation}
To derive the first of the Lanczos equations, we set $f(\lambda)=P_n(\lambda)^2$ in Eq.~\eqref{eq:symmetrized-saddle}. This gives
\begin{equation}\label{eq:first-string-monic}
\begin{aligned}
&\int d\lambda\,
V'(\lambda)\rho_0(\lambda)P_n(\lambda)^2
=
\int d\lambda\,dE\,
\rho_0(\lambda)\rho_0(E)
\frac{P_n(\lambda)^2-P_n(E)^2}{\lambda-E}
\\
=&
\int d\lambda\,dE\,
\rho_0(\lambda)\rho_0(E)
\left[
P_n(\lambda)
\frac{P_n(\lambda)-P_n(E)}{\lambda-E}
+
P_n(E)
\frac{P_n(\lambda)-P_n(E)}{\lambda-E}
\right].
\end{aligned}
\end{equation}
Using the difference of powers identity
\begin{equation*}
\frac{\lambda^j-E^j}{\lambda-E}
=
\sum_{\ell=0}^{j-1}\lambda^{j-1-\ell}E^\ell,
\end{equation*}
the difference quotient $[P_n(\lambda)-P_n(E)]/(\lambda-E)$ has degree at most $n-1$ in either variable. Orthogonality therefore makes both terms in the final line of Eq.~\eqref{eq:first-string-monic} vanish, and hence
\begin{equation}\label{eq:first-string}
\langle n|V'(\hat{\lambda})|n\rangle
=
\frac{1}{h_n}
\int d\lambda\,
V'(\lambda)\rho_0(\lambda)P_n(\lambda)^2
=0.
\end{equation}
For the second of the Lanczos equations, we choose $f(\lambda)=P_n(\lambda)P_{n-1}(\lambda)$. Eq.~\eqref{eq:symmetrized-saddle} then gives
\begin{equation}
\begin{aligned}
&\int d\lambda\,
V'(\lambda)\rho_0(\lambda)P_n(\lambda)P_{n-1}(\lambda)
\\
=&
\int d\lambda\,dE\,
\rho_0(\lambda)\rho_0(E)
\frac{P_n(\lambda)P_{n-1}(\lambda)-P_n(E)P_{n-1}(E)}{\lambda-E}
\\
=&
\int d\lambda\,dE\,
\rho_0(\lambda)\rho_0(E)
\left[
P_{n-1}(\lambda)
\frac{P_n(\lambda)-P_n(E)}{\lambda-E}
+
P_n(E)
\frac{P_{n-1}(\lambda)-P_{n-1}(E)}{\lambda-E}
\right].
\end{aligned}
\end{equation}
The second term vanishes because $[P_{n-1}(\lambda)-P_{n-1}(E)]/(\lambda-E)$ has degree at most $n-2$ in $E$. Since $P_n$ is monic, the remaining difference quotient can be expanded in the orthogonal polynomial basis as
\begin{equation}\label{eq:difference-quotient-expansion}
\frac{P_n(\lambda)-P_n(E)}{\lambda-E}
=
P_{n-1}(\lambda)
+
\sum_{k=0}^{n-2}c_k(E)P_k(\lambda).
\end{equation}
Orthogonality then gives the second of the Lanczos equations,
\begin{equation}\label{eq:second-string}
\begin{aligned}
\sqrt{R_n}\,
\langle n-1|V'(\hat{\lambda})|n\rangle
&=
\frac{1}{h_{n-1}}
\int d\lambda\,
V'(\lambda)\rho_0(\lambda)P_n(\lambda)P_{n-1}(\lambda)
\\
&=
\frac{1}{h_{n-1}}
\int d\lambda\,dE\,
\rho_0(\lambda)\rho_0(E)P_{n-1}(\lambda)^2
=1.
\end{aligned}
\end{equation}
Consequently, the recursion coefficients $R_n$ and $S_n$ satisfy the Lanczos equations
\begin{equation}\label{VSRREEQ2}
\begin{aligned}
\langle n|V'(\hat{\lambda})|n\rangle
&=0,
\\
\sqrt{R_n}\,\langle n-1|V'(\hat{\lambda})|n\rangle
&=1.
\end{aligned}
\end{equation}
Although the Lanczos equations~\eqref{VSRREEQ2} and the discrete string equations~\eqref{VSRREEQ} have the same structure on their left-hand sides, they generally determine different recursion coefficients. In particular, the right-hand side of the second relation is $1$ in Eq.~\eqref{VSRREEQ2}, whereas it is $n/N$ in Eq.~\eqref{VSRREEQ}. The corresponding polynomial families $P_n$ and $p_n$ are orthogonal with respect to the measures $d\mu(\lambda)=\rho_0(\lambda)d\lambda$ and $d\nu(\lambda)=e^{-NV(\lambda)}d\lambda$, respectively. The Coulomb-gas equation~\eqref{coulombgas2} relates the leading density of states $\rho_0(\lambda)$ to the potential $V(\lambda)$ and thereby accounts for the common structure of the two sets of equations. Under the normalizability assumptions~\eqref{boundarycondition} and for one-cut random matrix model, the recursion coefficients approach constants as $n\to\infty$, with $R_n\simeq R_{n+1}\simeq R_\infty$ and $S_n\simeq S_{n+1}\simeq S_\infty$. In this limit, the Lanczos equations~\eqref{VSRREEQ2} reduce to
\begin{equation}\label{extremaofthepotential}
\begin{aligned}
\frac{\partial}{\partial S_{\infty}}
\int_{0}^{\pi}
V\left(S_{\infty}+2\sqrt{R_{\infty}}\cos\theta\right)
\,d\theta&=0,\\
\frac{\partial}{\partial R_{\infty}}
\int_{0}^{\pi}
V\left(S_{\infty}+2\sqrt{R_{\infty}}\cos\theta\right)
\,d\theta&=\frac{\pi}{R_{\infty}}.
\end{aligned}
\end{equation}
Using the identification in Eq.~\eqref{coresplanrec}, this asymptotic behavior is equivalently expressed as
\begin{equation}
    \lim_{n\to\infty} a_n=a_{\infty}=S_\infty,
    \qquad
    \lim_{n\to\infty} b_n^2=b_{\infty}^2=R_\infty,
\end{equation}
where $S_\infty$ and $R_\infty$ are determined by Eq.~\eqref{extremaofthepotential}. To make the connection between this asymptotic behavior and the late-time growth of the spread complexity explicit, we introduce the spread complexity operator
\begin{equation}\label{krylovcomoprt}
\hat{C}
=
\sum_{n=0}^{\infty}n|n\rangle\langle n|,
\end{equation}
whose expectation value, $C(t)=\langle\psi(t)|\hat{C}|\psi(t)\rangle$, reproduces the definition in Eq.~\eqref{spreadcomplexity}. In the Heisenberg picture, this operator evolves according to
\begin{equation}
\frac{d\hat{C}}{dt}
=
i[\hat{\lambda},\hat{C}],
\end{equation}
Combining this equation with Eqs.~\eqref{compactrecursionrelation-rho0}, \eqref{coresplanrec}, and \eqref{krylovcomoprt} gives the acceleration operator
\begin{equation}\label{akrylovcom}
\begin{aligned}
\frac{d^2\hat{C}}{dt^2}
&=
i\left[\hat{\lambda},\frac{d\hat{C}}{dt}\right]
=
-\bigl[\hat{\lambda},[\hat{\lambda},\hat{C}]\bigr]
\\
&=
2\sum_{n=0}^{\infty}
\left(b_{n+1}^2-b_n^2\right)|n\rangle\langle n|
\\
&\quad+
\sum_{n=0}^{\infty}
b_{n+1}(a_{n+1}-a_n)
\left(
|n+1\rangle\langle n|
+
|n\rangle\langle n+1|
\right),
\end{aligned}
\end{equation}
This operator identity is the Ehrenfest theorem for spread complexity \cite{Erdmenger:2023wjg}. As the wave packet propagates toward large $n$ along the semi-infinite Krylov chain, the asymptotic constancy of $a_n$ and $b_n$ makes both $a_{n+1}-a_n$ and $b_{n+1}^2-b_n^2$ vanish, so that
\begin{equation}
\lim_{t\to\infty}\frac{d^2C}{dt^2}
=
\lim_{t\to\infty}
\left\langle\psi(t)\left|
\frac{d^2\hat{C}}{dt^2}
\right|\psi(t)\right\rangle
=0.
\end{equation}
Consequently, $C(t)$ becomes asymptotically linear, in agreement with the expected late-time behavior of holographic complexity \cite{Susskind:2014rva,Stanford:2014jda,
Brown:2015bva,Brown:2015lvg,Cai:2016xho,Guo:2017rul,Pedraza:2021mkh,Pedraza:2021fgp,Belin:2021bga,Belin:2022xmt,Pedraza:2022dqi,
Carrasco:2023fcj,Jorstad:2023kmq,Jiang:2023jti,Caceres:2023ziv,
Myers:2024vve,Arean:2024pzo,Jiang:2025qai,Miyaji:2025jxy,
Caceres:2025myu,Caceres:2025ypk,Fatemiabhari:2025cyy,
Fatemiabhari:2025usn,Fatemiabhari:2025poq,Nunez:2026vhw}. Having established the asymptotically linear behavior of $C(t)$, we now determine its late-time growth rate. Once the wave packet reaches the large-$n$ region, where the Lanczos coefficients approach $a_\infty$ and $b_\infty$, the Krylov amplitudes obey the discrete Schr\"odinger equation~\cite{Balasubramanian:2022tpr}
\begin{equation}\label{discsch2}
i\,\partial_t\psi_n(t)
=
b_\infty\,\psi_{n+1}(t)
+
a_\infty\,\psi_n(t)
+
b_\infty\,\psi_{n-1}(t).
\end{equation}
Translational invariance far from the boundary allows us to use the Bloch-wave ansatz
\begin{equation}
\psi_n(k,t)
=
e^{ikn-i\lambda(k)t},
\end{equation}
which yields the dispersion relation
\begin{equation}
\lambda(k)
=
a_\infty+2b_\infty\cos k.
\end{equation}
Since the spread complexity is the mean position of the wave packet along the Krylov chain, its asymptotic growth rate is obtained by averaging the group velocity towards increasing $n$ over the spectral density\footnote{We emphasize that this argument does not constitute a rigorous derivation of the late-time growth rate from the definition~\eqref{spreadcomplexity}. Rather, it provides a physical picture supported by numerical evidence. Establishing Eq.~\eqref{groupvelocity} rigorously from the definition of spread complexity remains an interesting problem for future work.},
\begin{equation}\label{groupvelocity}
\begin{aligned}
\lim_{t\to\infty}\frac{dC}{dt}
&=
\int d\lambda\,
\rho_0(\lambda)
\left|\frac{d\lambda(k)}{dk}\right|
\\
&=
\int d\lambda\,
\rho_0(\lambda)
\sqrt{4b_\infty^2-(\lambda-a_\infty)^2}
\leq
2b_\infty.
\end{aligned}
\end{equation}
The absolute value selects propagation towards increasing Krylov index. The
upper bound follows from the normalization of $\rho_0(\lambda)$ and is saturated only when all spectral weight is concentrated at the center of the band, $\rho_0(\lambda)=\delta(\lambda-a_\infty)$. The derivation above applies directly to one-cut ensembles, whose Lanczos coefficients approach
constants. In appendix~\ref{twocutrandom}, we further consider a two-cut matrix
model. Its Lanczos coefficients approach a period-two sequence, and the
resulting asymptotic Krylov chain still admits a Bloch description. We verify
the linear late-time growth and evaluate its rate using the first equality in
Eq.~\eqref{groupvelocity}.

\section{Geodesic distance interpretation}\label{sec:geodesic-distance}
To borrow a familiar aphorism, theoretical physics may not repeat itself, but it often rhymes. The recursion coefficients $R_n$ and $S_n$ governed by the Lanczos equations~\eqref{VSRREEQ2} provide a concrete example: in a seemingly unrelated combinatorial setting, they serve as generating functions for planar maps\footnote{Although $R_n$ and $S_n$ were previously shown to satisfy the Lanczos equations~\cite{Bouttier:2003dh,DiFrancesco:2004qj,di2005geodesic}, those derivations were purely combinatorial. Here, instead, we derive these equations from the leading density of states and construct the associated orthogonal-polynomial representation.} \cite{Bouttier:2003dh,DiFrancesco:2004qj,di2005geodesic}. A planar map is a connected graph embedded in the plane without edge crossings, considered up to topological equivalence. In the double-line expansion of a matrix integral, these maps appear as genus-zero ribbon graphs, or fatgraphs, and represent the planar Feynman diagrams of the model \cite{DiFrancesco:2004qj}.

\begin{figure}[!htbp]
\centering
\subfigure[]{
  \includegraphics[width=0.22\textwidth]{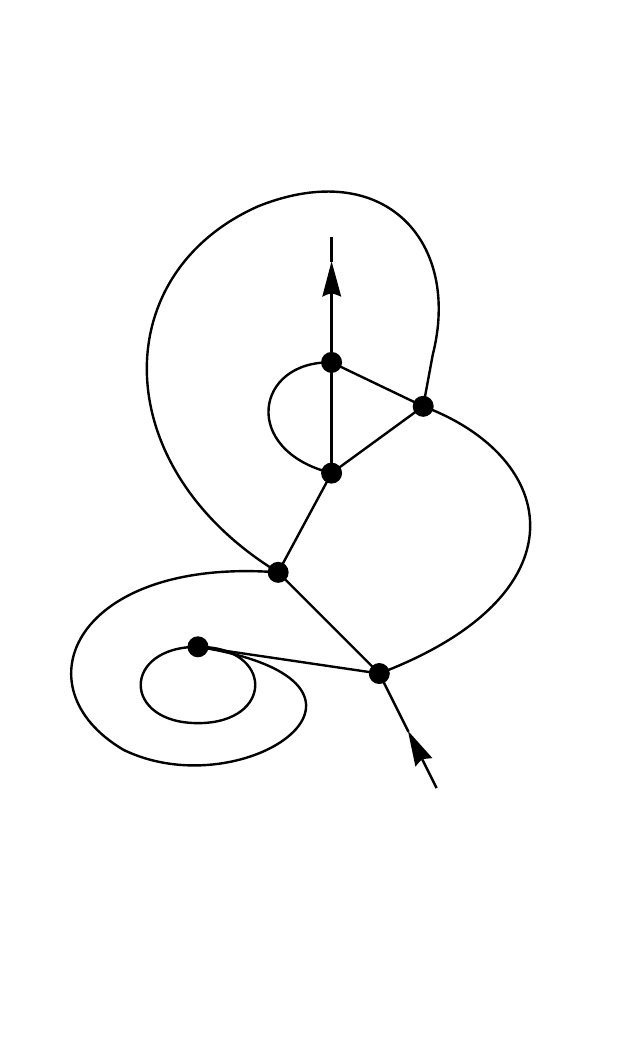}}
\hfill
\subfigure[]{
  \includegraphics[width=0.22\textwidth]{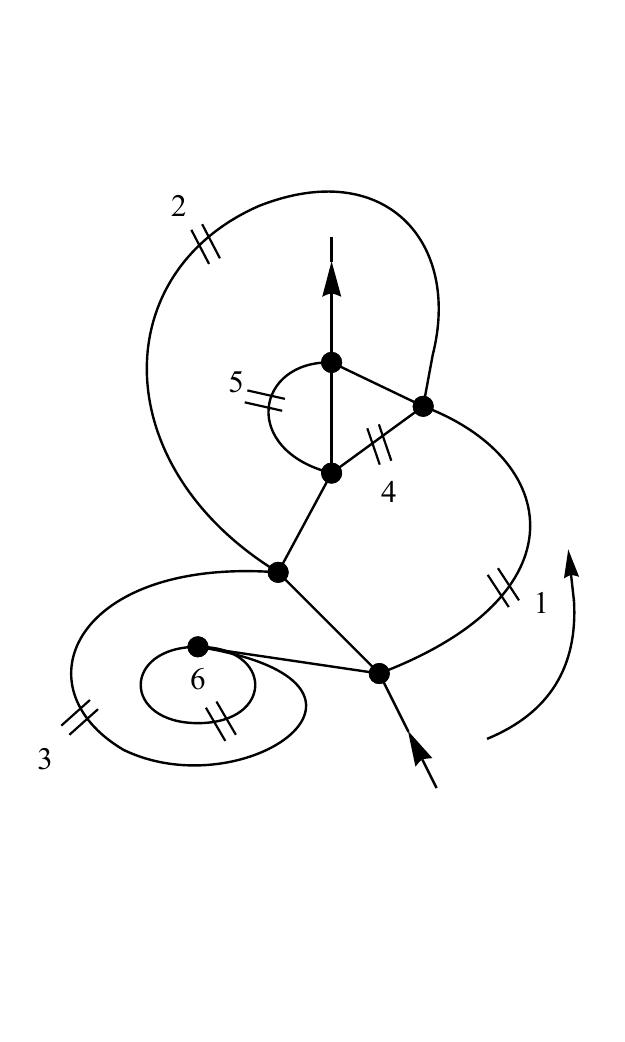}}
\hfill
\subfigure[]{
  \includegraphics[width=0.22\textwidth]{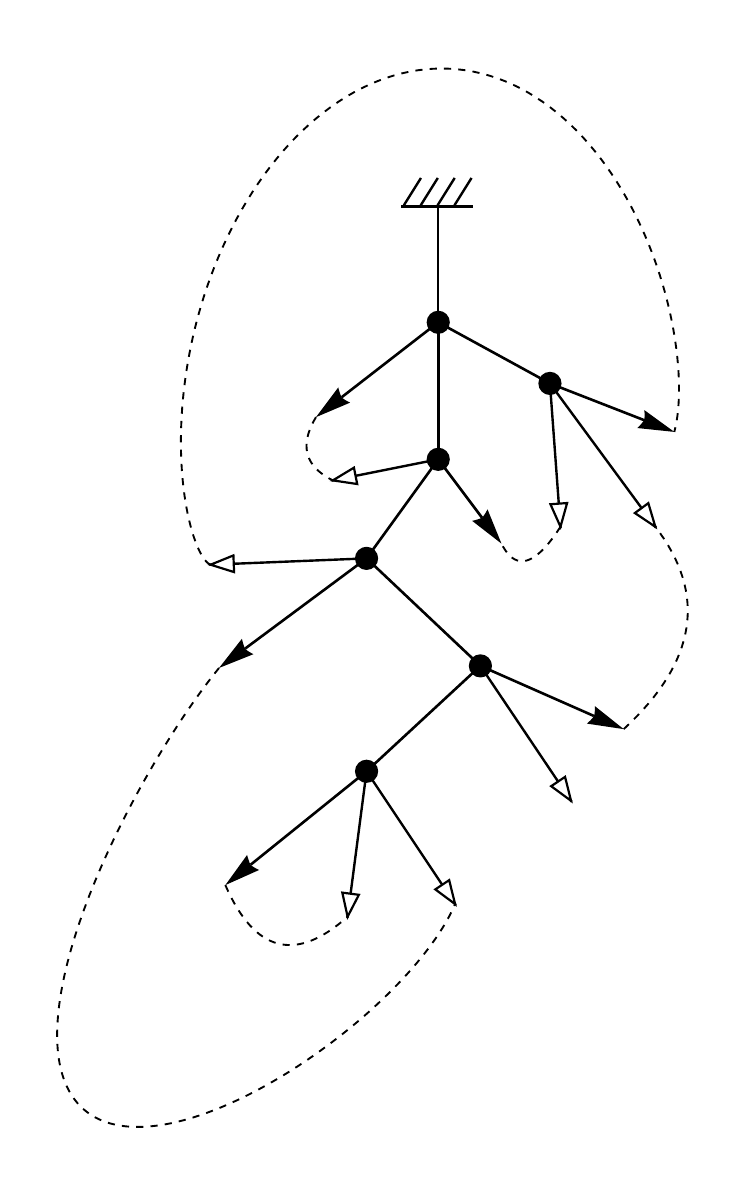}}
\hfill
\subfigure[]{
  \includegraphics[width=0.22\textwidth]{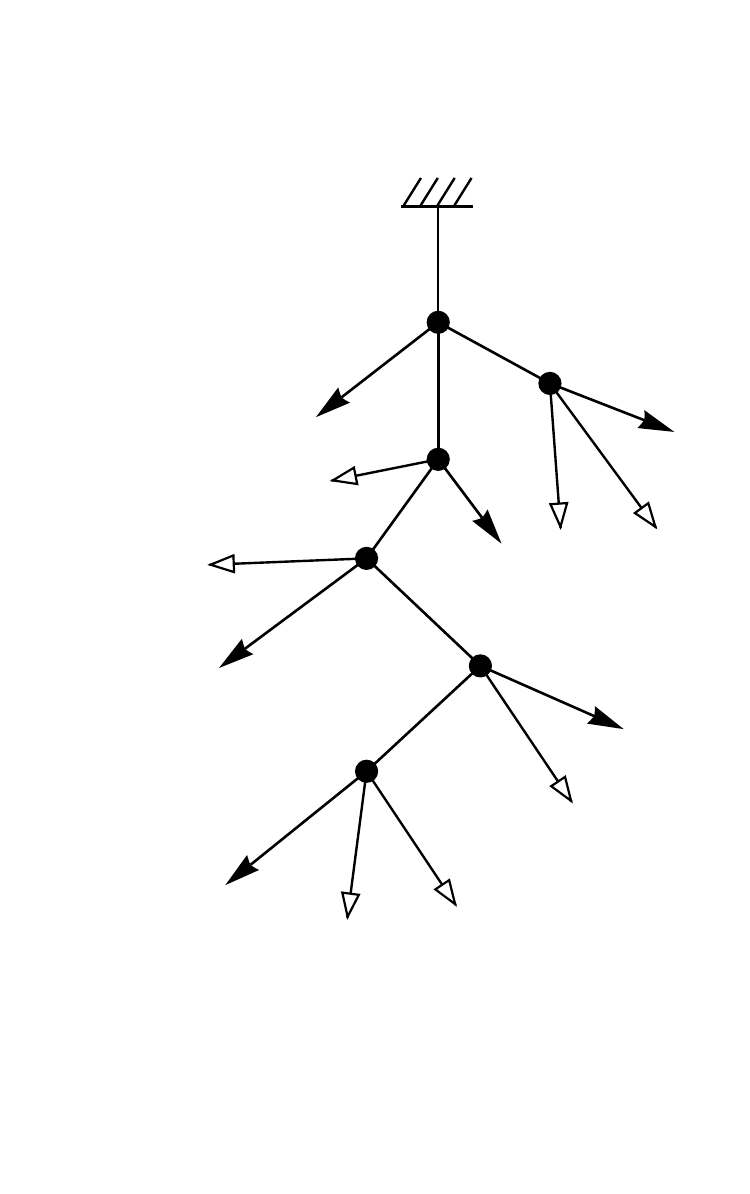}}
  \caption{Construction of the blossom tree associated with a planar map with two legs and four-valent internal vertices, reproduced from Ref.~\cite{Bouttier:2003dh}. Panel (a) displays the initial map. The numbers in panel (b) give the order of the allowed cuts: $1$--$3$ occur during the first circuit, namely the first complete counterclockwise traversal of the external boundary, and $4$--$6$ during the second. Each cut replaces an edge by a black bud and a white leaf, producing the intermediate object in panel (c). Turning the incoming leg into a leaf and the outgoing leg into the root gives the tree in panel (d). Pairing the buds and leaves in the reverse order reconstructs the original planar map.}
  \label{blossomtrees}
\end{figure}

For an even potential, $V(-\lambda)=V(\lambda)$, only even-valent internal vertices occur and $S_n=0$. In this case, $R_{n+1}-R_n$ is the generating function for planar maps with two legs whose marked legs are separated by the exact geodesic distance $n$ \cite{Bouttier:2003dh}. To explain this result, we review the relevant bijection between such maps and rooted blossom trees. A map with two legs carries two distinguished external legs, referred to as incoming and outgoing. A plane tree is an acyclic connected planar graph whose embedding fixes the ordering of its branches. A rooted blossom tree is a plane tree equipped with a distinguished root and two types of endpoints called buds and leaves. In Fig.~\ref{blossomtrees}, the buds and leaves are shown as black and white endpoints, respectively.

The forward map is obtained by following the boundary of the current external face counterclockwise, starting at the incoming leg. Whenever the traversal encounters an edge whose removal leaves the graph connected, that edge is cut and its two exposed half-edges are replaced by a bud and a leaf. In graph-theoretic language, a bridge is an edge whose deletion disconnects the graph, so the algorithm cuts precisely the edges that are not bridges encountered along the boundary. Repeating the traversal removes all cycles and leaves a plane tree. The incoming leg is then replaced by a leaf, and the outgoing leg is designated as the root.

The inverse map glues the tree back into a planar map with two legs. Each bud is iteratively paired with the closest available leaf in the counterclockwise direction, and each matched bud-leaf pair is fused into an edge. After all possible pairs have been glued, one leaf adjacent to the external face remains unmatched. This leaf becomes the incoming leg, while the root becomes the outgoing leg. The cutting and gluing procedures are inverse to one another, thereby establishing the required bijection \cite{schaeffer1997bijective,bouttier2002census}.

An important property of this bijection is that it preserves the geodesic distance between the two marked legs. For a planar map with two legs, this distance is defined as the minimum number of map edges crossed by a curve connecting the incoming leg to the outgoing leg. The corresponding quantity for the blossom tree is the depth of the root: after the pairs of buds and leaves are glued in the inverse construction, it is the number of nested edges separating the root from the external face. The cutting algorithm ensures that this root depth equals the geodesic distance of the original map \cite{schaeffer1997bijective,bouttier2002census}. Consequently, a map whose marked legs are separated by a distance $n$ is mapped to a blossom tree of root depth $n$; in the example shown in Fig.~\ref{blossomtrees}, both quantities are equal to one.

Let $G_n$ denote the generating function for blossom trees of root depth $n$, or equivalently for maps with two legs whose marked legs are separated by the exact geodesic distance $n$. Using the contour-walk representation, Bouttier, Di Francesco, and Guitter identified the root depth with the maximum height attained by the associated walk and showed that the exact-depth generating function is obtained by taking the difference between two consecutive cumulative generating functions \cite{Bouttier:2003dh}. Remarkably, in our conventions, these cumulative functions are precisely the recursion coefficients $R_n$ that obey the Lanczos equations~\eqref{VSRREEQ2}. Their result therefore becomes the relation,\footnote{In Ref.~\cite{Bouttier:2003dh}, the corresponding relation is written as $G_n=\widetilde{R}_n-\widetilde{R}_{n-1}$. Our index convention is shifted by one unit, $\widetilde{R}_n=R_{n+1}$, so the boundary condition $\widetilde{R}_{-1}=0$ corresponds to $R_0=0$.}
\begin{equation}\label{geodesicdifference}
G_n
=
R_{n+1}-R_{n}
=
b_{n+1}^2-b_n^2,
\end{equation}
where the last equality follows from Eq.~\eqref{coresplanrec}. For an even potential, $a_n=S_n=0$, so the off-diagonal term in the acceleration operator in Eq.~\eqref{akrylovcom} vanishes. Eq.~\eqref{geodesicdifference} then converts the remaining diagonal term into
\begin{equation}\label{geodesicoperatorrelation}
\frac{d^2\hat{C}}{dt^2}
=
2\sum_{n=0}^{\infty}
\left(b_{n+1}^2-b_n^2\right)|n\rangle\langle n|
=
2\sum_{n=0}^{\infty}G_n|n\rangle\langle n|
=
2\hat{G}.
\end{equation}
Here $\hat{G}=\sum_{n=0}^{\infty}G_n|n\rangle\langle n|$ is the geodesic generating operator: its eigenvalue on the Krylov basis $|n\rangle$ is the generating function for maps at the corresponding exact distance. Thus, the acceleration of the spread complexity is twice the expectation value of this combinatorial operator,
\begin{equation}\label{geodesicexpectationrelation}
\frac{d^2C(t)}{dt^2}
=
2\langle\psi(t)|\hat{G}|\psi(t)\rangle,
\end{equation}
This identity directly links propagation in Krylov space to geodesic-distance
enumeration in planar maps. By itself, however, this relation neither defines
a metric on the space of quantum states nor identifies $C(t)$ with a geodesic
length in Nielsen geometry or with a holographic complexity measure.
Nevertheless, recent work has argued that, for a class of one-cut random
matrix models with even potentials, the semiclassical Krylov Hilbert space may
be interpreted as the bulk Hilbert space of a dual gravitational
description~\cite{Bhattacharya:2026yfm}. Understanding the bulk interpretation
of $\ddot C(t)$ in this setting is an interesting direction for future work.

\section{Quartic potential}\label{sec:quartic-potential}
\begin{figure}[!htbp]
\centering
\subfigure[]{
  \includegraphics[width=0.7\textwidth]{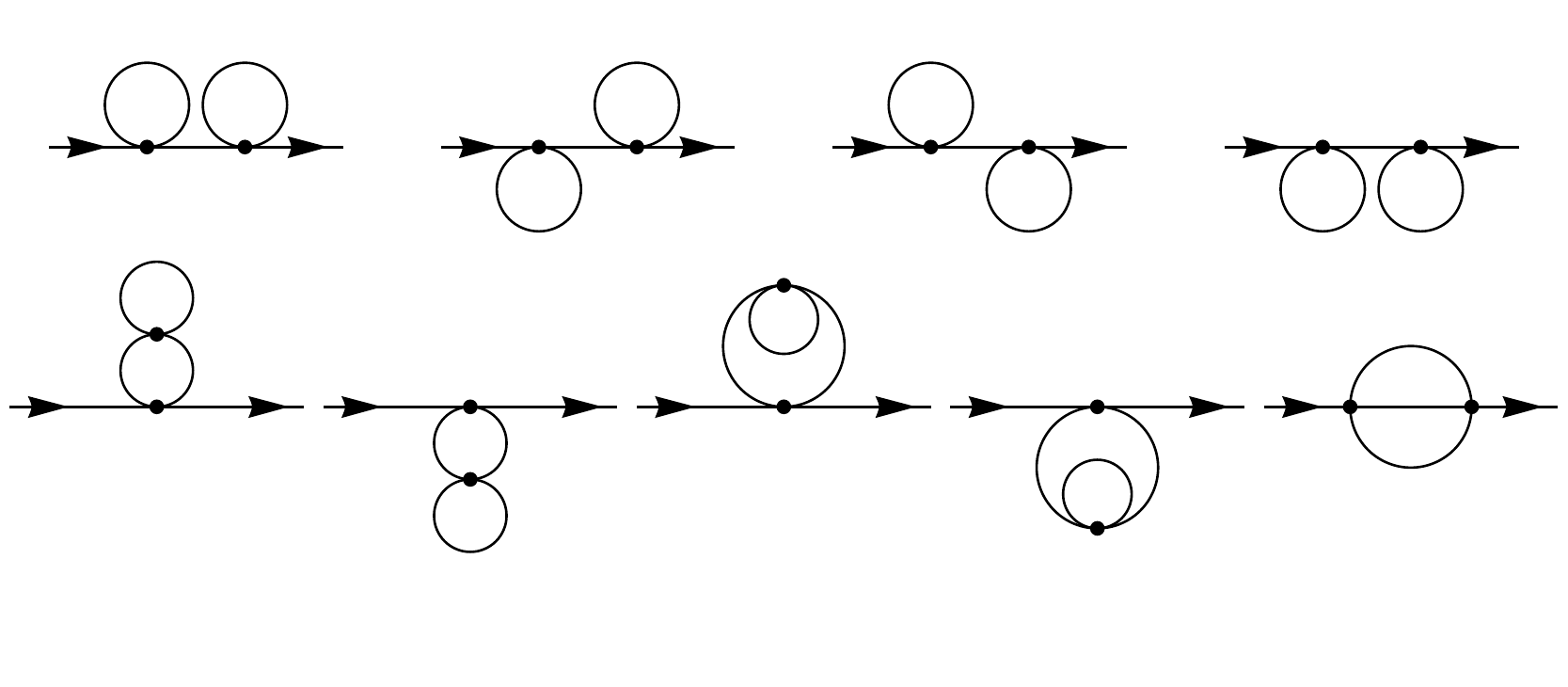}}
\par\vspace{-3mm}
\subfigure[]{
  \includegraphics[width=0.7\textwidth]{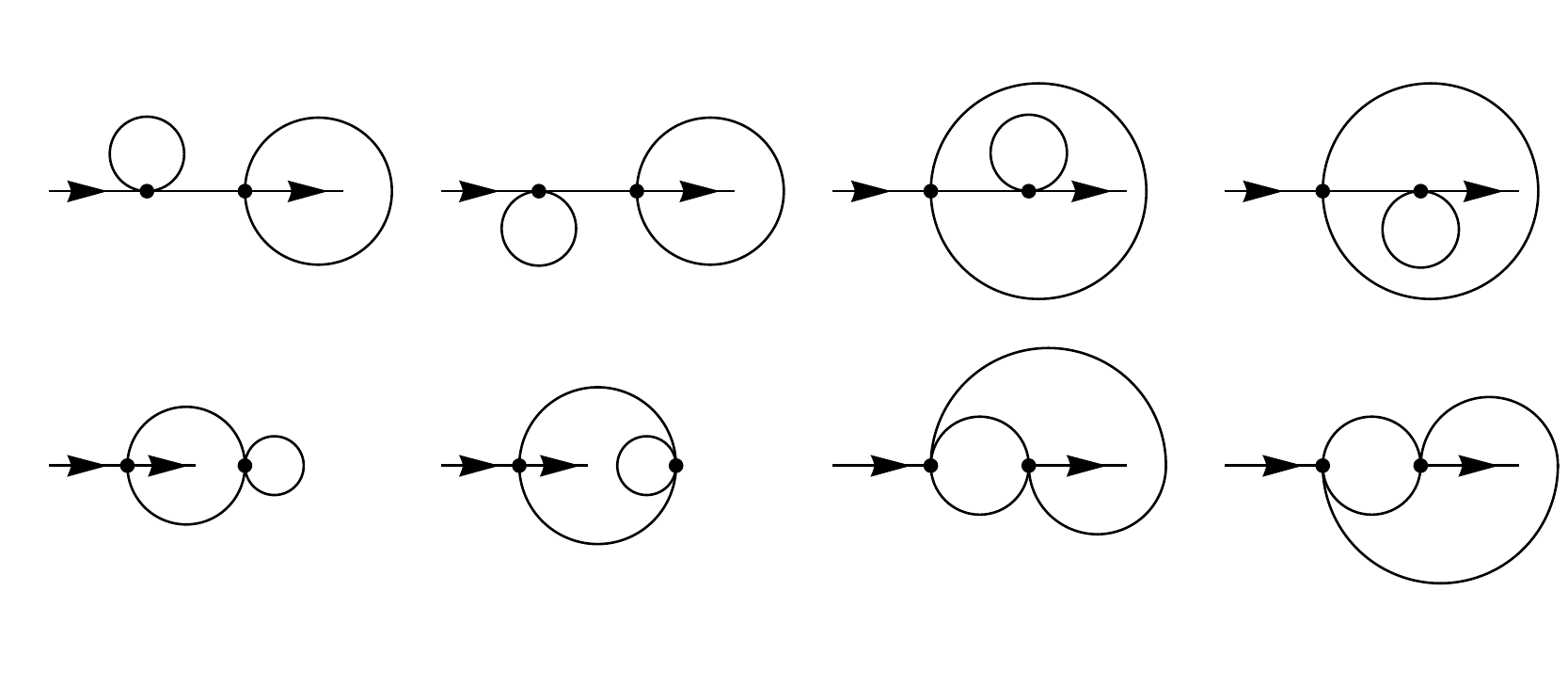}}
\par\vspace{-3mm}
\subfigure[]{
  \includegraphics[width=0.7\textwidth]{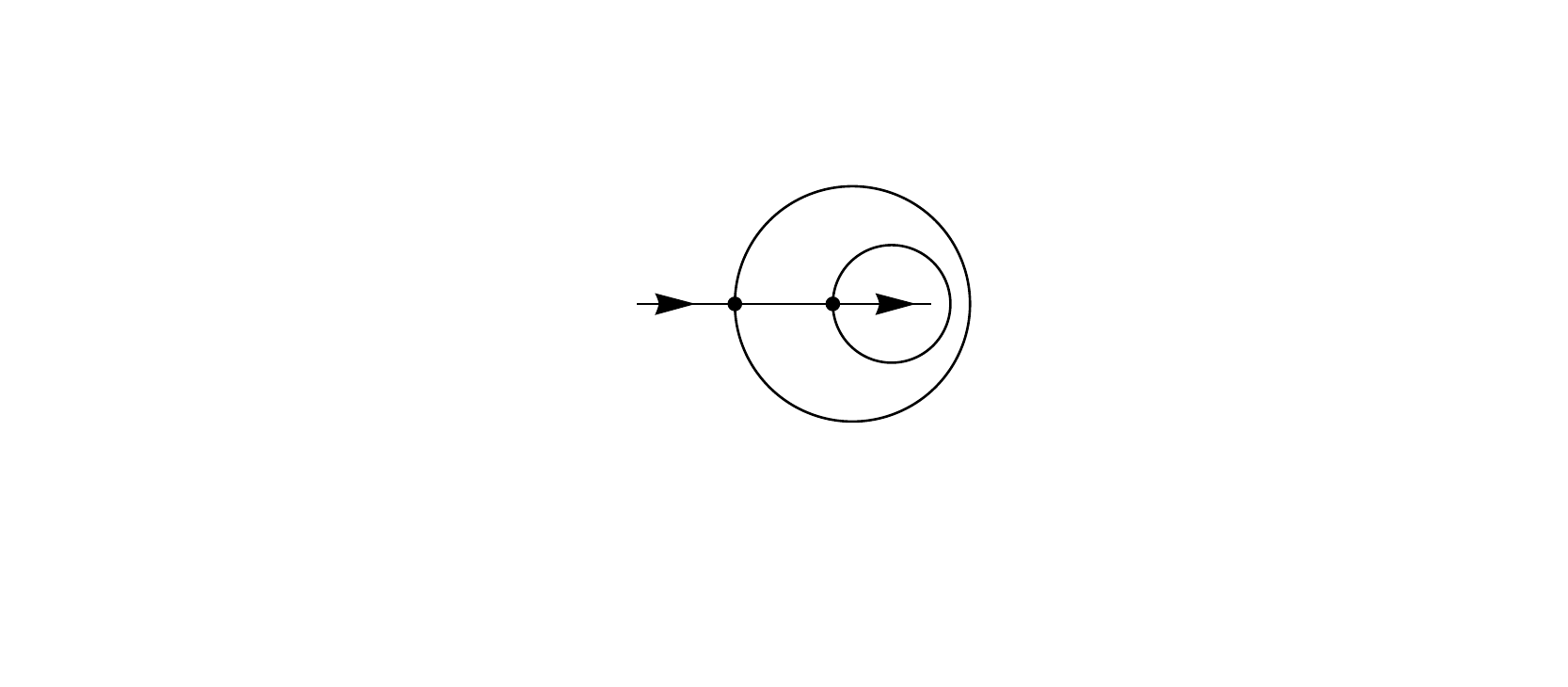}}
\par\vspace{-3mm}
\caption{Connected planar maps with two legs and two tetravalent internal vertices, reproduced from Ref.~\cite{Bouttier:2003dh}. The maps are classified by the geodesic distance between their marked legs: $n=0$ in panel (a), $n=1$ in panel (b), and $n=2$ in panel (c).}
\label{recurrencecoefficientquamain232}
\end{figure}

We now illustrate the preceding framework using the quartic potential
\begin{equation}
V_q(\lambda)
=
\frac{\lambda^2}{2}
-
\frac{g\lambda^4}{4},
\end{equation}
where we restrict to $g<0$, so that the potential is confining and the matrix
integral is convergent, while each internal four-valent vertex carries a factor
of $g$ in the associated formal combinatorial expansion. Because $V_q$ is even, the leading density of states has symmetric support on a single interval, which we write as $[-2\sqrt{R},2\sqrt{R}]$. Specializing the Coulomb-gas equation~\eqref{coulombgas-saddle} to $V_q$ gives
\begin{equation}\label{coulombgas3}
\frac{1}{2}V_q'(\lambda)
=
\int dE\,
\frac{\rho_0(E)}{\lambda-E},
\qquad
|\lambda|<2\sqrt{R}.
\end{equation}
Its solution with support on a single interval is
\begin{equation}\label{leadingdensstatesquar}
\rho_0(E)
=
\frac{1}{2\pi}
\left(
1-2gR-gE^2
\right)
\sqrt{4R-E^2},
\qquad
|E|\le 2\sqrt{R}.
\end{equation}
Here $\rho_0(E)=0$ outside the stated interval. For brevity, we write
$R=R_\infty$ for the
asymptotic recursion coefficient, which is determined by
\begin{equation}
R
=
1+3gR^2,
\qquad
R
=
\frac{1-\sqrt{1-12g}}{6g},
\qquad
R\longrightarrow 1
\quad\text{as}\quad
g\longrightarrow 0.
\end{equation}
The square-root singularity of $R(g)$ occurs at $g_c=1/12$. Since the density in Eq.~\eqref{leadingdensstatesquar} is even in $E$, all odd moments vanish. Using Eq.~\eqref{largeoments}, the even moments are
\begin{equation}
\lim_{N\to\infty}\langle m_{2n}\rangle
=
\int_{-2\sqrt{R}}^{2\sqrt{R}}
dE\,E^{2n}\rho_0(E)
=
\frac{R^n}{n+1}\binom{2n}{n}
\left(
1-\frac{3ngR^2}{n+2}
\right).
\end{equation}
The evenness of the density also implies $a_n=0$. Applying the recursive algorithm in Eq.~\eqref{recursivealgorithm} to the moments above yields the off-diagonal Lanczos coefficients
\begin{equation}\label{lanczoscoefficients-quartic}
b_n
=
\sqrt{
R\,
\frac{(1-y^n)(1-y^{n+3})}
{(1-y^{n+1})(1-y^{n+2})}
},
\qquad
n\geq 1.
\end{equation}
The auxiliary parameter $y$ is the branch with $|y|<1$ determined by
\begin{equation}
y+\frac{1}{y}+4
=
\frac{1}{gR},
\end{equation}
and satisfies $y\to0$ as $g\to0$. The result in Eq.~\eqref{lanczoscoefficients-quartic} also solves the quartic specialization of the Lanczos equations~\eqref{VSRREEQ2},
\begin{equation}
b_n^2
\left[
1-g\left(
b_{n-1}^2+b_n^2+b_{n+1}^2
\right)
\right]
=1,
\qquad
n\geq1.
\end{equation}
Here we used the correspondence $R_n=b_n^2$ from Eq.~\eqref{coresplanrec}. As established in Eq.~\eqref{geodesicdifference}, the generating function for maps with two legs at exact geodesic distance $n$ is
\begin{equation}
G_n(g)
=
b_{n+1}^2-b_n^2,
\qquad
n\geq0.
\end{equation}
The first three generating functions are \cite{Bouttier:2003dh}
\begin{equation}
\label{twolegdiagrams}
\begin{aligned}
G_0(g)
&=
\frac{
  24g-1+\sqrt{1-12g}
}{
  9g\bigl(1+\sqrt{1-12g}\bigr)
}
=
1+2g+9g^2+54g^3+378g^4+\cdots,
\\[0.5em]
G_1(g)
&=
\frac{
  27g-2+(2-15g)\sqrt{1-12g}
}{
  27g\bigl(1-8g+\sqrt{1-12g}\bigr)
}
=
g+8g^2+65g^3+554g^4+\cdots,
\\[0.5em]
G_2(g)
&=
\frac{
  252g^2-6g-1+(1+12g)\sqrt{1-12g}
}{
  27\Bigl[
    190g^2-51g+3
    +(3-33g+44g^2)\sqrt{1-12g}
  \Bigr]
}
=
g^2+15g^3+\cdots.
\end{aligned}
\end{equation}
The coefficient of $g^k$ in $G_n(g)$ counts maps with two legs, $k$ tetravalent internal vertices, and geodesic distance $n$. Figure~\ref{recurrencecoefficientquamain232} displays all such maps with $k=2$: nine have distance zero, eight have distance one, and one has distance two, in agreement with the coefficients of $g^2$ in Eq.~\eqref{twolegdiagrams}.
We next consider the monic polynomials orthogonal with respect to $d\mu(E)=\rho_0(E)\,dE$. For the density in Eq.~\eqref{leadingdensstatesquar}, they can be expressed in terms of Rogers polynomials, also known as continuous $q$-ultraspherical polynomials,
\begin{equation}\label{Rogerspolynomials}
P_n(E)
=
R^{n/2}
\frac{(y;y)_n}{(y^2;y)_n}
C_n\left(
    \frac{E}{2\sqrt{R}};
    y^2\,\middle|\,y
  \right),
\end{equation}
where $C_n(x;\beta\mid q)$ denotes the Rogers polynomial of degree $n$. We use the finite and infinite $q$-Pochhammer symbols
\begin{equation}
(a;q)_n
=
\prod_{k=0}^{n-1}(1-aq^k),
\qquad
(a;q)_\infty
=
\prod_{k=0}^{\infty}(1-aq^k),
\end{equation}
with $(a;q)_0=1$. The prefactor in Eq.~\eqref{Rogerspolynomials} makes $P_n(E)$ monic. After the change of variables $E=2\sqrt{R}\cos\theta$, the orthogonality relation in Eq.~\eqref{innerproduct-rho0} becomes
\begin{equation}
\begin{aligned}
&\int_{-2\sqrt{R}}^{2\sqrt{R}}
dE\,\rho_0(E)P_n(E)P_m(E)
\\
=&
R^{(n+m)/2}
  \frac{(1-y)^3}
{2\pi(1-y^3)(1-y^{n+1})(1-y^{m+1})}
\\
&\times
\int_0^\pi
  d\theta\,
  \left|
\frac{(e^{2\mathrm{i}\theta};y)_\infty}
{(y^2e^{2\mathrm{i}\theta};y)_\infty}
\right|^2
  C_n(\cos\theta;y^2\mid y)
  C_m(\cos\theta;y^2\mid y)
\\
=&
h_n\delta_{nm},
\end{aligned}
\end{equation}
where the squared norm is
\begin{equation}
h_n
=
\prod_{j=1}^{n}b_j^2
=
R^n
\frac{(1-y)(1-y^{n+3})}
{(1-y^{n+1})(1-y^3)}.
\end{equation}
In the final step, we used the standard orthogonality relation for the Rogers polynomials,
\begin{equation}
\begin{aligned}
&\frac{1}{2\pi}
\int_{0}^{\pi}
d\theta\,\left|
\frac{(e^{2\mathrm{i}\theta};y)_\infty}
{(y^2e^{2\mathrm{i}\theta};y)_\infty}
\right|^2
C_n(\cos\theta;\beta\mid q)
C_m(\cos\theta;\beta\mid q)
\\
=&
\frac{(\beta;q)_\infty(\beta q;q)_\infty}
{(\beta^2;q)_\infty(q;q)_\infty}
\frac{(1-\beta)(\beta^2;q)_n}
{(1-\beta q^n)(q;q)_n}
\delta_{nm}.
\end{aligned}
\end{equation}
This formula holds for $|q|<1$ and $|\beta|<1$; here it is applied with $q=y$ and $\beta=y^2$.
In the Gaussian limit $g\to0$, one has $R\to1$, $y\to0$, and $b_n\to1$ for every $n\geq1$. The Rogers polynomials in Eq.~\eqref{Rogerspolynomials} then reduce to the Chebyshev polynomials of the second kind,
\begin{equation}
\lim_{g\to0}P_n(E)
=
U_n\!\left(\frac{E}{2}\right),
\end{equation}
where $U_n(x)$ denotes the Chebyshev polynomial of the second kind with degree $n$. 
\begin{figure}[!htbp]
\centering
\subfigure[$C(t)$ as a function of time]{
  \includegraphics[width=0.45\textwidth]{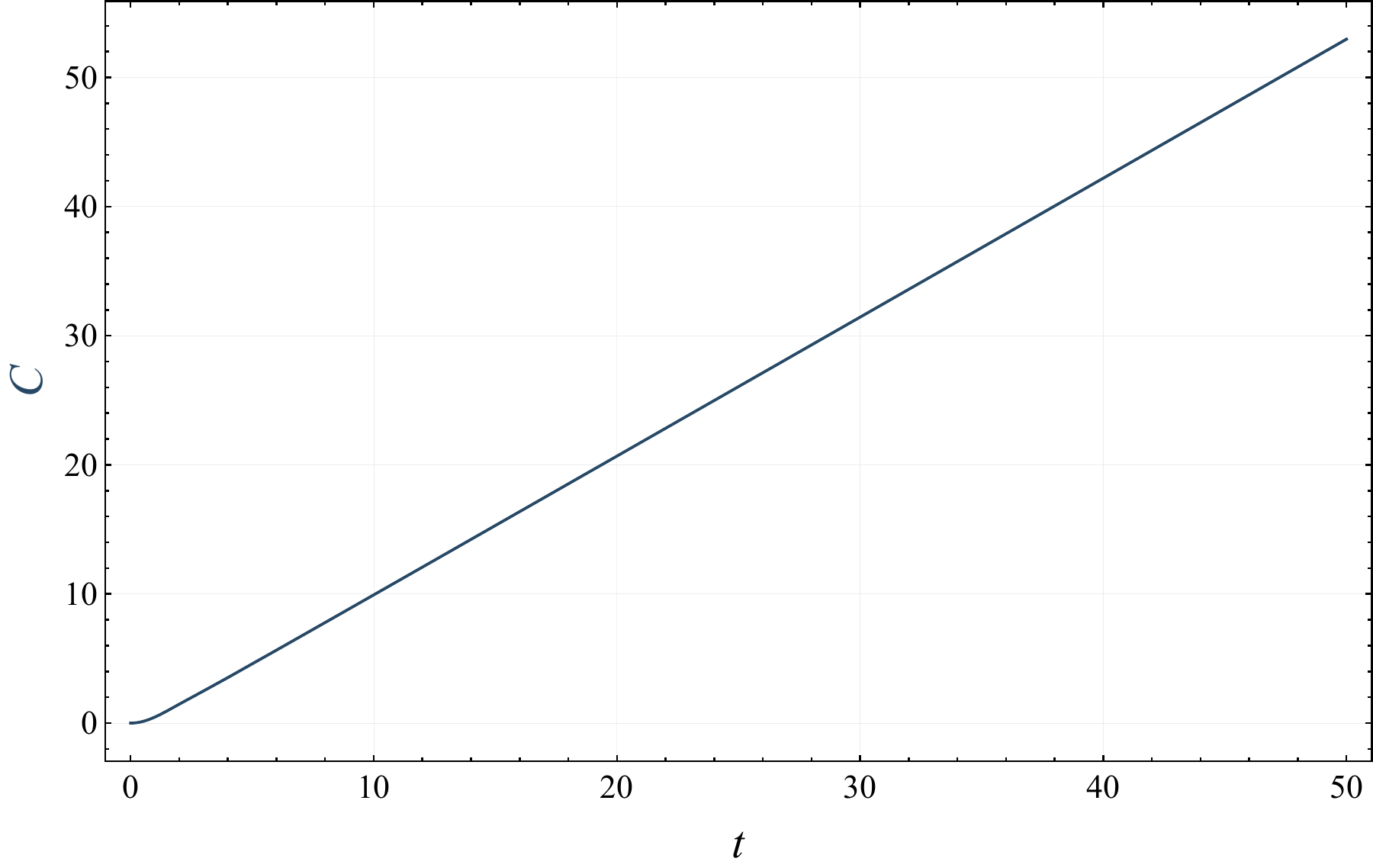}}
\hfill
\subfigure[$\dot{C}(t)$ as a function of time]{
  \includegraphics[width=0.45\textwidth]{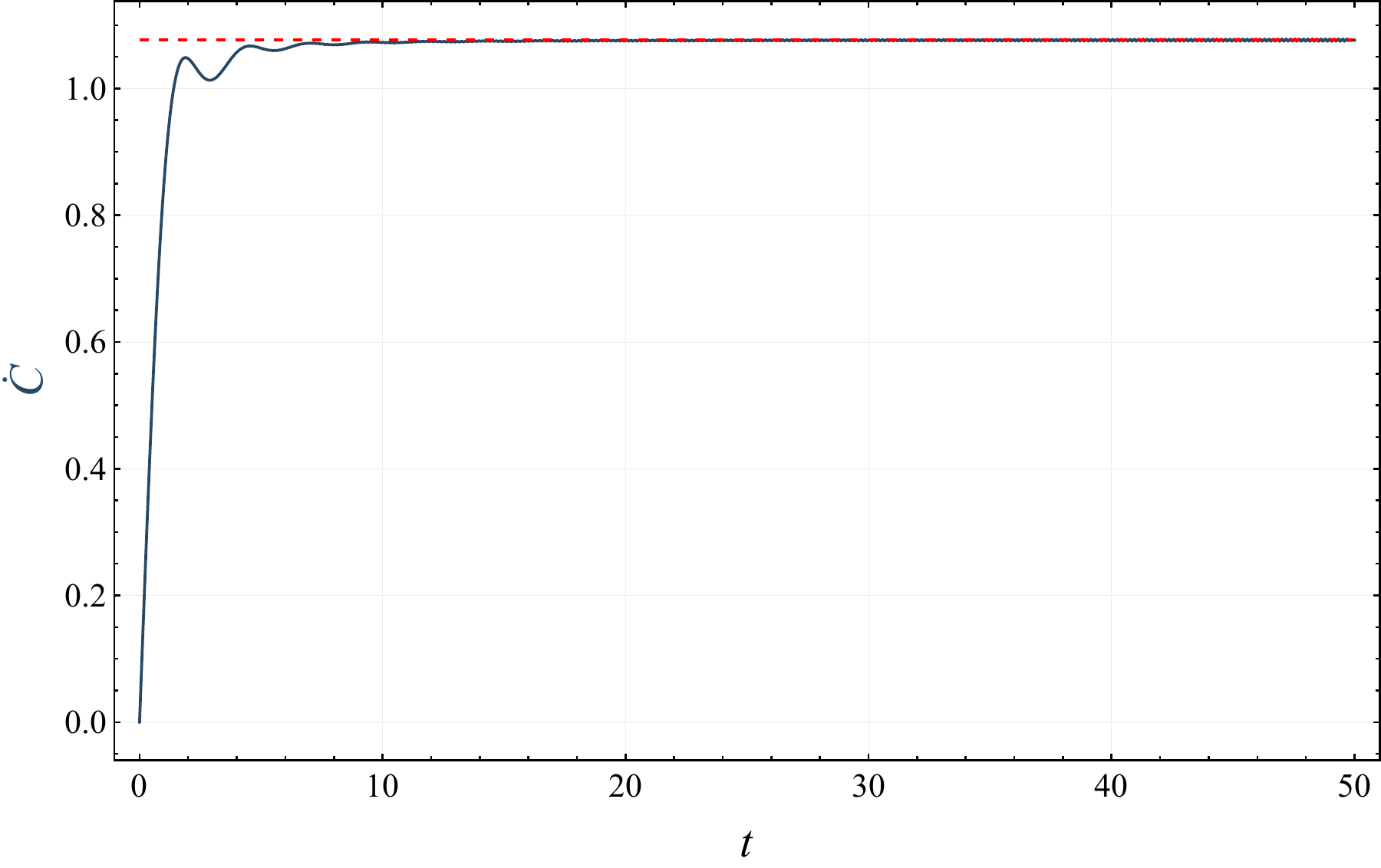}}
\caption{Spread complexity $C(t)$ and its growth rate $\dot C(t)$ for the quartic model at $g=-1$. In panel (b), the red dashed line indicates the
analytic late-time growth rate given in Eq.~\eqref{latetimegrowthratequ}.}
\label{fig:Quartickkrylovcomplexity}
\end{figure}
Since $h_n=1$ in this limit, substituting this result into Eq.~\eqref{spreadcomplexity} gives \cite{Barbon:2019wsy,Rabinovici:2023yex,Balasubramanian:2025xkj}
\begin{equation}
\begin{aligned}
C(t)
&=
\sum_{n=0}^{\infty}n
\int d\mu(E)\,d\mu(E')\,
U_n\!\left(\frac{E}{2}\right)
U_n\!\left(\frac{E'}{2}\right)
e^{-\mathrm{i}(E-E')t}
\\
&=
\frac{1}{t^2}
\sum_{n=1}^{\infty}
n(n+1)^2J_{n+1}^2(2t),
\end{aligned}
\end{equation}
where $J_n(x)$ is the Bessel function of the first kind of integer order $n$. Writing $\dot C(t)=dC(t)/dt$, the late-time growth rate approaches the constant
\begin{equation}\label{cdotinfit}
\lim_{t\to\infty}\dot{C}(t)
=
\lim_{t\to\infty}\frac{2}{t^2}
\sum_{m=1}^{\infty}
m(m+1)
J_m(2t)J_{m+1}(2t)
=
\frac{16}{3\pi}.
\end{equation}
For general $g<0$ in the one-cut regime, substituting the density in
Eq.~\eqref{leadingdensstatesquar} into Eq.~\eqref{groupvelocity}, with
$a_\infty=0$ and $b_\infty^2=R(g)$, gives the late-time growth rate
\begin{equation}\label{latetimegrowthratequ}
\begin{aligned}
\lim_{t\to\infty}\frac{dC}{dt}
&=
\int d\lambda\,
\rho_0(\lambda)
\sqrt{4b_\infty^2-\lambda^2}
\\
&=
\frac{16}{3\pi}
R(g)^{3/2}
\left(1-\frac{14}{5}gR(g)\right).
\end{aligned}
\end{equation}
In the Gaussian limit $g\to0$, where $R(g)\to1$, this result reduces to
Eq.~\eqref{cdotinfit}. At the representative confining coupling $g=-1$,
Fig.~\ref{fig:Quartickkrylovcomplexity} shows that $C(t)$ crosses over from
quadratic to linear growth and that $\dot C(t)$ approaches the analytic value
in Eq.~\eqref{latetimegrowthratequ}.

\section{Double-scaled SYK model}\label{sec:dssyk}
The Sachdev--Ye--Kitaev (SYK) model consists of $M$ Majorana fermions with
random all-to-all interactions involving $p$ fermions~\cite{Maldacena:2016hyu}. Its
Hamiltonian is
\begin{equation}\label{dssyk-hamiltonian}
H
=
i^{p/2}
\sum_{1\leq i_1<\cdots<i_p\leq M}
J_{i_1\cdots i_p}\,
\psi_{i_1}\cdots\psi_{i_p},
\end{equation}
where $p$ is even, the Majorana operators $\psi_i$ obey the usual
anticommutation relations, and the couplings $J_{i_1\cdots i_p}$ are
independent Gaussian random variables with zero mean and a variance fixed by
the choice of energy scale. The double-scaled SYK (DSSYK) limit is defined by
taking $p,M\to\infty$ while holding
$q=e^{-2p^2/M}\in(0,1)$
fixed~\cite{Cotler:2016fpe,Garcia-Garcia:2018fns,Berkooz:2018jqr}. In this
limit, the normalized leading density of states is the $q$-normal
distribution \cite{Sachdev:1992fk,Berkooz:2018qkz}
\begin{equation}\label{dosDSSYK}
\rho_0(E)
=
\frac{\sqrt{1-q}}
{\pi\sqrt{1-\frac{1-q}{4}E^2}}
\prod_{k=0}^{\infty}
\left(
\frac{1-q^{2k+2}}
{1-q^{2k+1}}
\right)
\left(
1-
\frac{(1-q)q^k}
{(1+q^k)^2}
E^2
\right), \qquad
|E|\le \frac{2}{\sqrt{1-q}}.
\end{equation}
The density vanishes outside this interval and interpolates between the
Wigner semicircle distribution as $q\to0$ and the normal distribution as
$q\to1$. Up to an irrelevant additive constant, the matrix potential whose
Coulomb-gas saddle reproduces Eq.~\eqref{dosDSSYK} has the exact Chebyshev
expansion~\cite{Jafferis:2022wez}
\begin{equation}\label{eq:q-matrix-potential}
V_s(\lambda)
=
\sum_{n=1}^{\infty}
\frac{(-1)^{n-1}}{n}
q^{n^2/2}
\left(
q^{n/2}+q^{-n/2}
\right)
T_{2n}
\!\left(
\frac{\sqrt{1-q}}{2}\lambda
\right).
\end{equation}
Here $T_n$ is the Chebyshev polynomial of the first kind, defined by
$T_n(\cos\theta)=\cos(n\theta)$. Since both the potential and the density are
even, all odd moments vanish, while Eq.~\eqref{largeoments} gives the even moments
as
\begin{equation}\label{dssyk-even-moments}
\lim_{N\to\infty}\langle m_{2n}\rangle
=
\int dE\,E^{2n}\rho_0(E)
=
\frac{1}{(1-q)^n}
\sum_{j=-n}^{n}
(-1)^j
q^{j(j-1)/2}
\binom{2n}{n+j}.
\end{equation}
The same expression follows from the chord diagram expansion, in which each
crossing of two chords contributes a factor of $q$~\cite{Berkooz:2018jqr}.
Applying the recursive algorithm in Eq.~\eqref{recursivealgorithm} then yields the
Lanczos coefficients~\cite{Nandy:2024zcd,Xu:2024gfm}
\begin{equation}\label{lanczoscoefficients-dssyk}
a_n
=
0,
\qquad
b_n
=
\sqrt{\frac{1-q^n}{1-q}},
\qquad n\geq1,
\end{equation}
with $b_0=0$. The vanishing diagonal coefficients $a_n$ reflect the symmetry
of the spectral measure.
We now construct the monic polynomials orthogonal with respect to
$d\mu(\lambda)=\rho_0(\lambda)\,d\lambda$. For the density in
Eq.~\eqref{dosDSSYK}, these polynomials are rescaled continuous $q$-Hermite
polynomials \cite{Lin:2022rbf,Rabinovici:2023yex},
\begin{equation}\label{eq:Pn-q-Hermite}
P_n(\lambda)
=
\frac{1}{(1-q)^{n/2}}
H_n\!\left(
\frac{\sqrt{1-q}}{2}\lambda
\,\middle|\,q
\right),
\end{equation}
where $H_n(x\mid q)$ denotes the continuous $q$-Hermite polynomial of degree
$n$. With the change of variables
$\lambda=2\cos\theta/\sqrt{1-q}$, the orthogonality relation in
Eq.~\eqref{innerproduct-rho0} becomes
\begin{equation}
\begin{aligned}
&\int_{-\frac{2}{\sqrt{1-q}}}^{\frac{2}{\sqrt{1-q}}}
d\lambda\,
\rho_0(\lambda)
P_n(\lambda)P_m(\lambda)
\\
=&
\frac{(q;q)_\infty}
{2\pi(1-q)^{(n+m)/2}}
\int_0^\pi
d\theta\,
\left|
\left(e^{2\mathrm{i}\theta};q\right)_\infty
\right|^2
H_n(\cos\theta\mid q)
H_m(\cos\theta\mid q)
\\
=&
h_n\delta_{nm},
\end{aligned}
\label{q-Hermite-orthogonality-rho0}
\end{equation}
where the squared norm is
\begin{equation}
h_n
=
\prod_{j=1}^{n}b_j^2
=
\frac{(q;q)_n}{(1-q)^n}.
\end{equation}
In the last equality, we used the standard orthogonality relation for the
continuous $q$-Hermite polynomials,
\begin{equation}
\begin{aligned}
&\frac{1}{2\pi}
\int_0^\pi
d\theta\,
\left|
\left(e^{2\mathrm{i}\theta};q\right)_\infty
\right|^2
H_n(\cos\theta\mid q)
H_m(\cos\theta\mid q)
=
\frac{(q;q)_n}{(q;q)_\infty}
\delta_{nm}.
\end{aligned}
\label{continuous-q-Hermite-orthogonality}
\end{equation}
It follows that the normalized polynomial states introduced in Eq.~\eqref{auxiliaryHilbertspace},
\begin{equation}
|n\rangle
=
\frac{P_n(\lambda)}{\sqrt{h_n}},
\qquad n=0,1,2,\ldots,
\end{equation}
obey the same Jacobi recurrence as the normalized states of fixed chord number
in DSSYK. 
\begin{figure}[!htbp]
\centering
\subfigure[$C(t)$ as a function of time]{
  \includegraphics[width=0.45\textwidth]{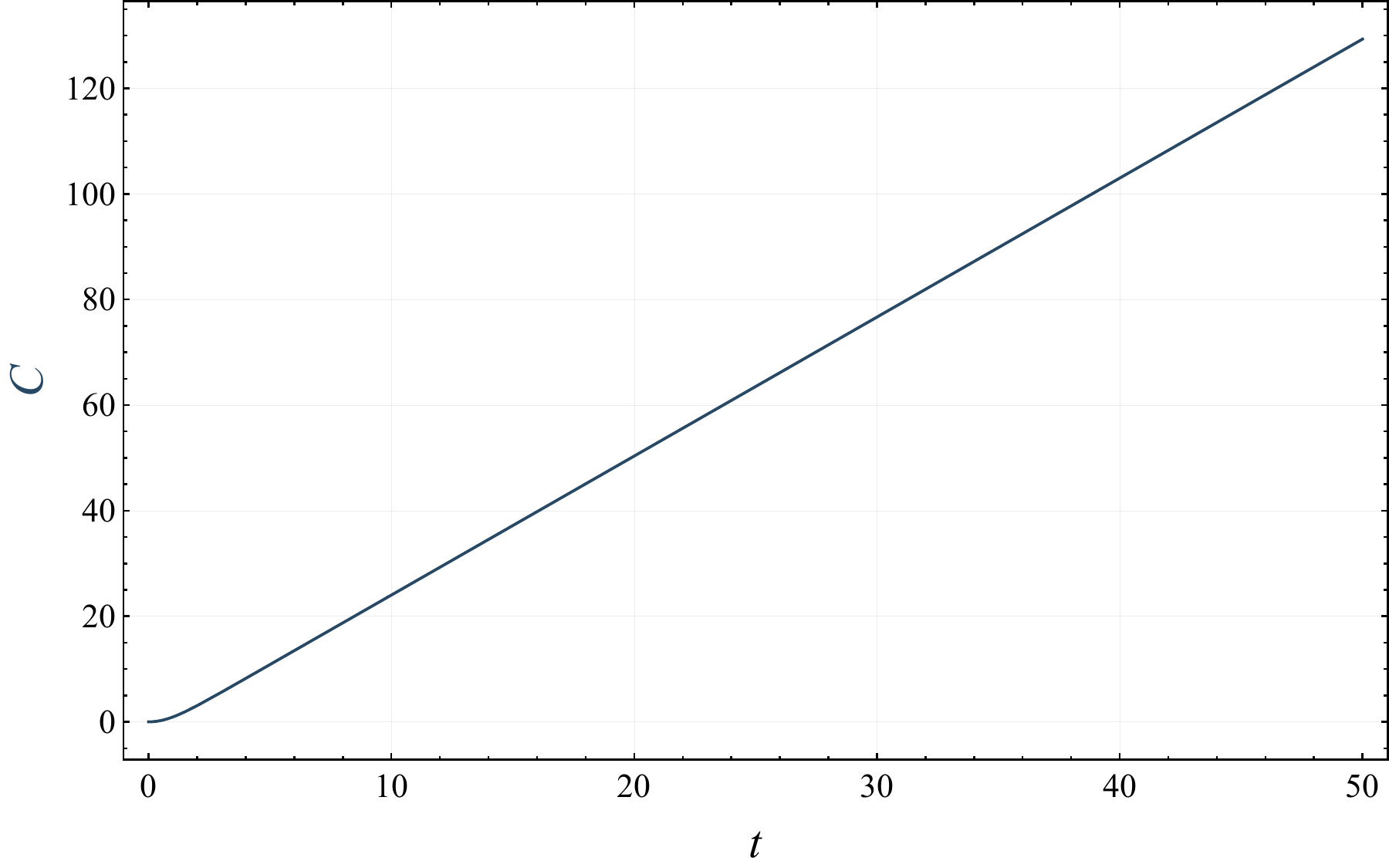}}
\hfill
\subfigure[$\dot{C}(t)$ as a function of time]{
  \includegraphics[width=0.45\textwidth]{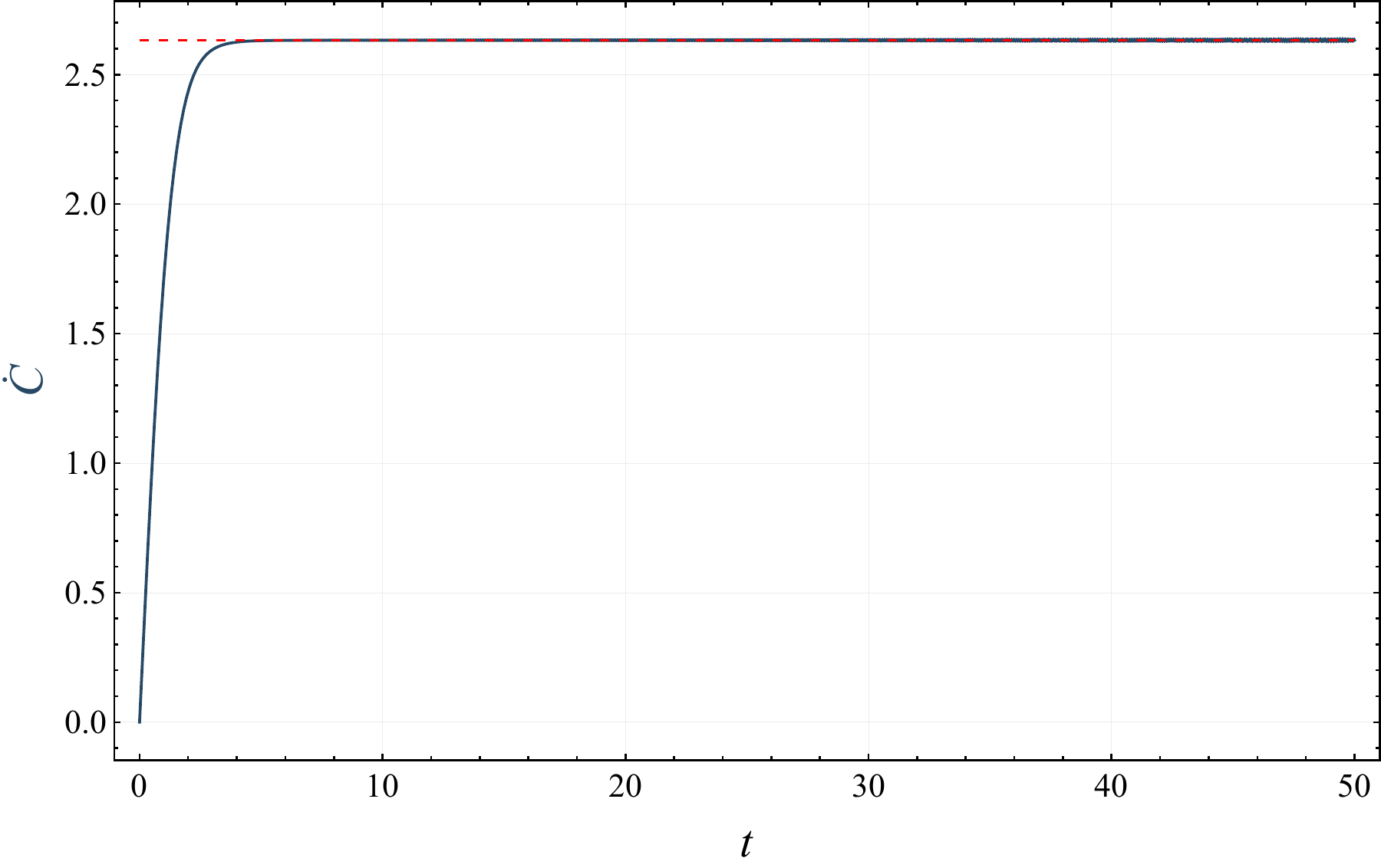}}
\caption{Spread complexity $C(t)$ and its growth rate $\dot C(t)$ in the
double-scaled SYK model at $q=0.5$. In panel (b), the red dashed line indicates the
analytic late-time growth rate given in Eq.~\eqref{latetimegrowthratedssyk}.}
\label{fig:DSSYKkrylovcomplexity}
\end{figure}
Under the bulk interpretation of the chord Hilbert space, the chord
number provides a discrete measure of the length of the two-sided
wormhole \cite{Lin:2022rbf,Rabinovici:2023yex}. An additional result
of the present work is that the coefficients in
Eq.~\eqref{lanczoscoefficients-dssyk} and the matrix potential in
Eq.~\eqref{eq:q-matrix-potential} satisfy the Lanczos equations in
Eq.~\eqref{VSRREEQ2}. Since $V(\lambda)$ is even, the first relation in
Eq.~\eqref{VSRREEQ2} follows immediately from parity, whereas the nontrivial
equation is
\begin{equation}\label{lanczos-equation-dssyk}
b_n\,\langle n-1|V'(\hat{\lambda})|n\rangle
=1.
\end{equation}
Thus, for the matrix model defined by the potential \(V_s(\lambda)\), the difference between two consecutive squared Lanczos coefficients gives the generating function for maps with a fixed geodesic distance. Substituting the Lanczos coefficients in Eq.~\eqref{lanczoscoefficients-dssyk} into Eq.~\eqref{geodesicdifference}, we find that the generating function for two-legged maps with exact geodesic distance $n$ is
\begin{equation}
G_n
=
b_{n+1}^2-b_n^2=\frac{1-q^{n+1}}{1-q}-\frac{1-q^n}{1-q}=q^n.
\end{equation}
Although the DSSYK matrix potential in Eq.~\eqref{eq:q-matrix-potential}
contains interactions of arbitrarily high order, the resulting generating
function reduces to the remarkably simple form $q^n$. The correspondence
between planar maps and chord diagrams developed in Ref.~\cite{Budd:2026bfh}
suggests that this simplification may admit a direct explanation in terms of
geodesic chord diagrams. We leave this interesting question for future work. We now turn to the two endpoints of the interpolation. In the
$q\to0$ semicircle limit, $b_n\to1$ for every $n\geq1$, and the continuous
$q$-Hermite polynomials in Eq.~\eqref{eq:Pn-q-Hermite} reduce to the Chebyshev
polynomials of the second kind,
\begin{equation}
\lim_{q\to0}P_n(E)
=
U_n\!\left(\frac{E}{2}\right),
\end{equation}
which is precisely the $g=0$ case discussed in the previous section. In the opposite limit,
$q\to1$, one instead finds $b_n\to\sqrt{n}$ at fixed $n$.\footnote{The limits \(n\to\infty\) and \(q\to1\) do not commute. For fixed \(0\leq q<1\), the Lanczos coefficients approach \(b_n\to(1-q)^{-1/2}\) as \(n\to\infty\). By contrast, taking \(q\to1\) first gives \(b_n=\sqrt n\), which grows without bound. In the latter limit, the potential behaves at large \(|\lambda|\) as
\begin{equation}
    V_s(\lambda)
=
\log(\lambda^2)+\gamma+\log 2+O(\lambda^{-2}),
\end{equation}
where $\gamma \approx 0.57$ is Euler's constant. Consequently, \(V_s(\lambda)-\log(1+\lambda^2)\) approaches the finite constant \(\gamma+\log 2\), rather than \(+\infty\). The normalizability condition~\eqref{boundarycondition} is therefore violated, and the resulting asymptotic constancy of the Lanczos coefficients does not extend to the \(q\to1\) limit.} The continuous
$q$-Hermite polynomials correspondingly reduce to rescaled Hermite
polynomials,
\begin{equation}\label{PNHERM}
\lim_{q\to1}P_n(\lambda)
= 2^{-n/2}\,
H_n\!\left(\frac{\lambda}{\sqrt{2}}\right),
\qquad
H_n(\lambda)
= (-1)^n e^{\lambda^{2}}
  \frac{d^{\,n}}{d\lambda^{n}} e^{-\lambda^{2}} .
\end{equation}
Here $H_n(\lambda)$ is the Hermite polynomial. In this limit,
$d\mu(E)$ becomes the standard normal measure and $h_n=n!$; substituting
Eq.~\eqref{PNHERM} into Eq.~\eqref{spreadcomplexity} therefore gives
\begin{equation}
\begin{aligned}
C(t)
=
\sum_{n=0}^{\infty}\frac{2^{-n}n}{n!}
\int d\mu(E)\,d\mu(E')\,
H_n\!\left(\frac{E}{\sqrt{2}}\right)
H_n\!\left(\frac{E'}{\sqrt{2}}\right)
e^{-\mathrm{i}(E-E')t}
=t^2.
\end{aligned}
\end{equation}
Thus, at the endpoint $q=1$, the spread complexity remains quadratic rather
than becoming linear at late times because the limiting potential fails to
satisfy the normalizability condition~\eqref{boundarycondition}. By contrast, for fixed $0\leq q<1$, the Lanczos coefficients approach the constant
$(1-q)^{-1/2}$, and the preceding asymptotic argument implies linear
late-time growth. Substituting $a_\infty=0$, $b_\infty^2=(1-q)^{-1}$, and the density of states in
Eq.~\eqref{dosDSSYK} into Eq.~\eqref{groupvelocity}, we obtain the
late-time growth rate for fixed $0\leq q<1$,
\begin{equation}\label{latetimegrowthratedssyk}
\begin{aligned}
\lim_{t\to\infty}\frac{dC}{dt}
&=
\int d\lambda\,
\rho_0(\lambda)
\sqrt{4b_\infty^2-\lambda^2}
\\
&=
\frac{(q;q)_\infty}{\pi\sqrt{1-q}}
\int_0^\pi d\theta\,\sin\theta\,
\left|\left(e^{2\mathrm{i}\theta};q\right)_\infty\right|^2
\\
&=
\frac{4}{\pi\sqrt{1-q}}
\left[
1+\sum_{m=1}^{\infty}
\frac{(-1)^m q^{m(m-1)/2}(1+q^m)}
{1-4m^2}
\right].
\end{aligned}
\end{equation}
At $q=0$, this expression reduces to the Gaussian result in
Eq.~\eqref{cdotinfit}. For the representative value $q=0.5$,
Fig.~\ref{fig:DSSYKkrylovcomplexity} confirms the result numerically: $C(t)$ is
quadratic at early times and becomes linear as $\dot C(t)$ approaches the
analytic value in Eq.~\eqref{latetimegrowthratedssyk}.

\section{Conclusions and outlook}\label{Conclusion}
In this work, we studied the spread complexity of the
infinite-temperature thermofield double state in large-$N$ random matrix
theory. The leading density of states~\eqref{leadingdenstates} determines the survival amplitude and
its moments~\eqref{largeoments}, which in turn define the family of orthogonal polynomials~\eqref{innerproduct-rho0} whose
recursion coefficients are the Lanczos coefficients of the corresponding
Krylov dynamics. Starting from the Coulomb-gas equation~\eqref{coulombgas-saddle}, we
derived a pair of nonlinear relations for these coefficients, which we called
the Lanczos equations~\eqref{VSRREEQ2}. Although their left-hand sides have the same structure
as those of the discrete string equations~\eqref{VSRREEQ}, the two systems involve different
orthogonality measures and differ on the right-hand side. For one-cut
ensembles satisfying the normalizability condition~\eqref{boundarycondition},
the Lanczos coefficients approach constants at large $n$,
and the asymptotic Lanczos equations determine these constants directly from
the matrix potential. In the Ehrenfest relation~\eqref{akrylovcom}, this
convergence makes the spread complexity acceleration vanish at late times,
implying asymptotically linear growth. The resulting translationally invariant
Krylov chain admits a Bloch-wave description, from which the late-time growth rate
follows as the spectral average of the group velocity.

Our central result is a geometric and combinatorial interpretation of the
Lanczos coefficients. For an even potential, the difference
$G_n=b_{n+1}^2-b_n^2$ is the generating function for planar maps with two
marked legs separated by the exact geodesic distance $n$. This relation converts the acceleration operator into
$d^2\hat C/dt^2=2\hat G$, where $\hat G$ is the geodesic generating operator.
Consequently, the acceleration of the spread complexity is twice the
expectation value of $\hat G$ in the evolving state. This result provides a direct interpretation of the Lanczos coefficients beyond their conventional role as hopping amplitudes and on-site energies along the Krylov chain.

We illustrated the framework with the quartic matrix model and the
DSSYK model. In the quartic model, the coefficient
of $g^k$ in $b_{n+1}^2-b_n^2$ counts planar maps with two legs and $k$
tetravalent internal vertices and geodesic distance $n$. The Lanczos
coefficients are encoded by Rogers polynomials and approach a constant at
large $n$, while the numerical evolution exhibits quadratic early-time growth followed by linear late-time growth. In DSSYK, the
$q$-normal density leads to continuous $q$-Hermite polynomials and the coefficients $a_n=0$ and $b_n^2=(1-q^n)/(1-q)$. The auxiliary Hilbert space can then be identified with the chord Hilbert space spanned by states of fixed chord number. For fixed $0\leq q<1$, the Lanczos coefficients become constant at large
$n$ and the spread complexity grows linearly at late times. 

Several extensions would be worthwhile. For a potential that is not even, the
diagonal coefficients $a_n$ do not vanish, and the acceleration operator
contains off-diagonal terms. Extending the geodesic distance interpretation to this
case may reveal the combinatorial meaning of the full operator rather than
only its diagonal part. It would also be interesting to understand corrections at finite $N$,
for which maps of higher genus should supplement the planar description. The present
analysis is restricted to the infinite-temperature TFD state. Extending the Lanczos
equations to finite temperature and determining how temperature affects the spread
complexity would therefore provide a natural extension of the present framework.
Finally, double-scaled matrix
models provide nonperturbative descriptions of two-dimensional quantum
gravity~\cite{Banks:1989df,Seiberg:2004at,Saad:2019lba,Johnson:2019eik,Johnson:2020heh,Johnson:2020exp,
Johnson:2022wsr}, while in DSSYK the
chord number is related to the length of the two-sided
wormhole~\cite{Lin:2022rbf,Rabinovici:2023yex}. These observations suggest that the geodesic
generating operator may admit a bulk interpretation. Clarifying this
possibility and its relation to recent geometric descriptions of spread
complexity~\cite{Caputa:2024sux,Zhai:2024tkz} is a natural direction for future work.

\acknowledgments
I thank Pawel Caputa, Ben Craps, Yu-Xiao Liu, Juan F.~Pedraza and Shan-Ming Ruan for valuable correspondence. I acknowledge support from the Chinese Scholarship Council (CSC) through a graduate scholarship. I acknowledge support from the Spanish Agencia Estatal de Investigación through grants CEX2025-001574-S, PID2021-123017NB-I00 and PID2024-156043NB-I00, funded by MCIN/AEI/10.13039/501100011033, and ERDF, EU.

\appendix

\section{Two-cut random matrix model}
\label{twocutrandom}

We now consider a two-cut matrix model defined by the quartic double-well
potential
\begin{equation}
V_p(\lambda)
=
\frac{x}{2}\lambda^2
+
\frac{1}{4}\lambda^4,
\end{equation}
with $x<-2$. The leading density of states is supported on
$[-c,-d]\cup[d,c]$, where
\begin{equation}
c=\sqrt{2-x},
\qquad
d=\sqrt{-2-x},
\end{equation}
and the Coulomb-gas equation~\eqref{coulombgas-saddle} gives
\begin{equation}\label{leaddensittwo}
\rho_0(E)
=
\frac{1}{2\pi}|E|\sqrt{(c^2-E^2)(E^2-d^2)}.
\end{equation}
Because this density is even, $a_n=0$ and all odd moments vanish. The
remaining moments follow from Eq.~\eqref{largeoments}:
\begin{equation}
\lim_{N\to\infty}\langle m_{2n}\rangle
=
\int dE\,E^{2n}\rho_0(E)
=
\sum_{j=0}^{\lfloor n/2 \rfloor}
\frac{1}{j+1}
\binom{2j}{j}
\binom{n}{2j}
\left(\frac{c^2+d^2}{2}\right)^{n-2j}.
\end{equation}
Applying the recursion algorithm~\eqref{recursivealgorithm} then yields
\begin{equation}\label{lanczoscoefficients-two}
b_{2n}
=
\sqrt{
\frac{U_{n-1}\!\left(-\frac{x}{2}\right)}
{U_n\!\left(-\frac{x}{2}\right)}
},
\qquad
b_{2n+1}
=
\sqrt{
\frac{U_{n+1}\!\left(-\frac{x}{2}\right)}
{U_n\!\left(-\frac{x}{2}\right)}
}.
\end{equation}
These coefficients satisfy the quartic specialization of the Lanczos
equations~\eqref{VSRREEQ2},
\begin{equation}
b_n^2
\left(
x+
b_{n-1}^2+b_n^2+b_{n+1}^2
\right)
=1,
\qquad
n\geq1,
\end{equation}
where $R_n=b_n^2$ as in Eq.~\eqref{coresplanrec}. Their large-$n$ limit is
the period-two sequence
\begin{equation}
\lim_{n\to\infty} b_{2n}
=
\frac{c-d}{2},
\qquad
\lim_{n\to\infty} b_{2n+1}
=
\frac{c+d}{2}.
\label{eq:lanczos-period-two}
\end{equation}
The asymptotic chain is therefore the Su--Schrieffer--Heeger model, whose two
Bloch bands have the dispersion relation~\cite{asboth2015short}
\begin{equation}
\lambda_{\pm}(k)
=
\pm\sqrt{-x+2\cos k}.
\end{equation}
The two signs correspond to the two components of the spectral support. In
terms of the energy, the magnitude of the group velocity is
\begin{equation}
\left|\frac{d\lambda}{dk}\right|
=
\frac{\sqrt{(c^2-\lambda^2)(\lambda^2-d^2)}}{2|\lambda|}.
\end{equation}
\begin{figure}[!htbp]
\centering
\subfigure[$C(t)$]{
  \includegraphics[width=0.45\textwidth]{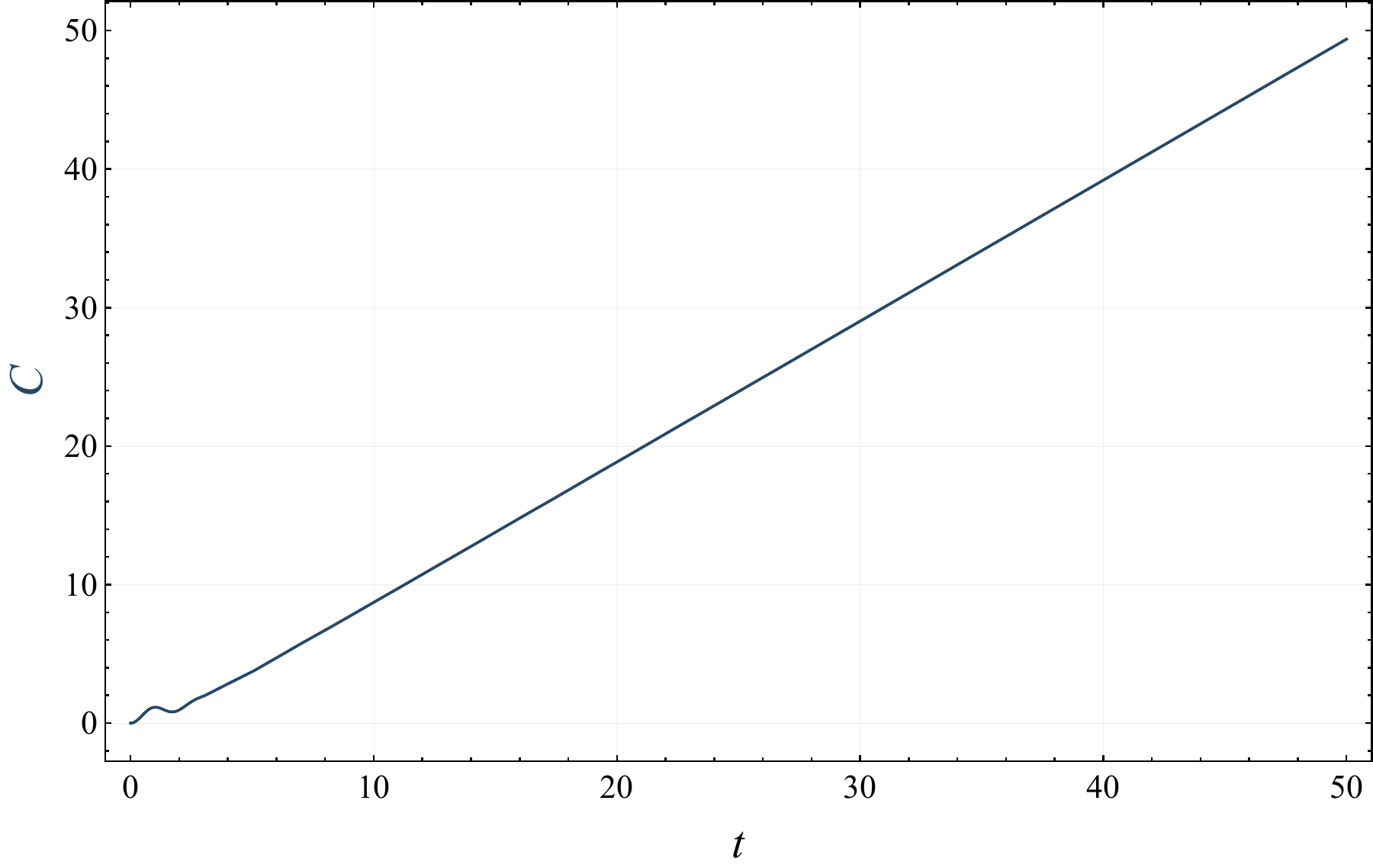}}
\hfill
\subfigure[$\dot{C}(t)$]{
  \includegraphics[width=0.45\textwidth]{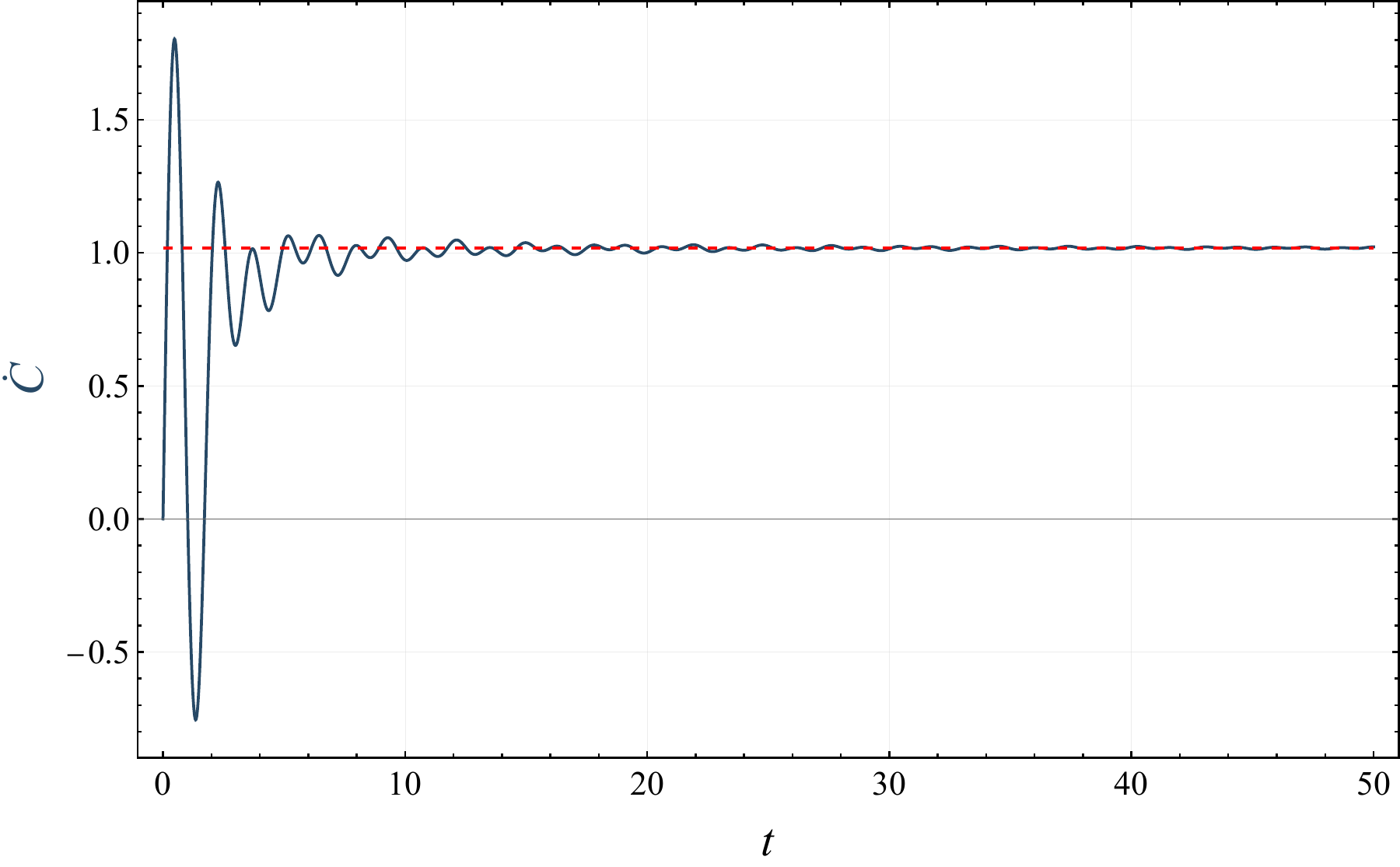}}
\caption{Spread complexity $C(t)$ and its growth rate $\dot C(t)$ in the
two-cut model at $x=-3$. In panel (b), the red dashed line indicates the
analytic late-time growth rate given in Eq.~\eqref{Thelategrowthtwocut}.}
\label{fig:two-cut-complexity}
\end{figure}
Since one Bloch unit cell contains two Krylov sites, the velocity average in
Eq.~\eqref{groupvelocity} acquires an overall factor of two. The late-time
growth rate is thus
\begin{equation}\label{Thelategrowthtwocut}
\begin{aligned}
\lim_{t\to\infty}\frac{dC}{dt}
&=
2\int d\lambda\,
\rho_0(\lambda)
\left|\frac{d\lambda}{dk}\right|
\\
&=
\frac{2}{15\pi}
(c-d)^3\left(c^2+3cd+d^2\right).
\end{aligned}
\end{equation}
This result provides an explicit two-cut realization of the linear growth
derived in section~\ref{sec:lanczos-equations}.
Figure~\ref{fig:two-cut-complexity} shows the spread complexity and its growth
rate for $x=-3$. The numerical growth rate approaches the analytic value in
Eq.~\eqref{Thelategrowthtwocut}, confirming the asymptotically linear growth of
$C(t)$. Motivated by this example, we conjecture that the same Bloch-wave
construction extends to general multi-cut models whenever the asymptotic
Lanczos coefficients define a periodic Jacobi operator, with the linear
late-time growth rate determined by the corresponding dispersion
relation~\cite{teschl2000jacobi}.

\newpage
\bibliography{Refs}
\bibliographystyle{JHEP}

\end{document}